\documentclass[10pt]{book}
\usepackage{array}
\usepackage{amsmath}
\usepackage{amssymb}
\usepackage{amsfonts}
\usepackage{hyperref}
\usepackage{amsthm}
\usepackage{latexsym}
\usepackage{graphicx}

\newtheorem{lemma}{Lemma}
\newtheorem{corollary}{Corollary}

\newtheorem{theorem}{Theorem}
\newtheorem{remark}{Remark}

\DeclareMathOperator{\var}{Var}

\begin{document}
\renewcommand{\baselinestretch}{1.2}
\markright{
}
\markboth{\hfill{\footnotesize\rm DAVID A. DEGRAS 
}\hfill}
{\hfill {\footnotesize\rm CONFIDENCE BANDS FOR FUNCTIONAL DATA} \hfill}
\renewcommand{\thefootnote}{}
$\ $\par
\fontsize{10.95}{14pt plus.8pt minus .6pt}\selectfont
\vspace{0.8pc}
\centerline{\large\bf SIMULTANEOUS CONFIDENCE BANDS FOR NONPARAMETRIC}
\vspace{2pt}
\centerline{\large\bf REGRESSION WITH FUNCTIONAL DATA}
\vspace{.4cm}
\centerline{David A. Degras}
\vspace{.4cm}
\centerline{\it University of Chicago}
\vspace{.55cm}
\fontsize{9}{11.5pt plus.8pt minus .6pt}\selectfont

\begin{quotation}
\noindent {\it Abstract:}
We consider nonparametric regression in the context of  
functional data, that is, 
when a random sample of functions is observed on a fine grid. 
We obtain a functional asymptotic normality result 
allowing to build  simultaneous confidence bands (SCB) for various 
 estimation and inference tasks. 
Two applications to a SCB procedure for the regression function 
and to a goodness-of-fit test for curvilinear regression models are proposed.   
The first one has improved accuracy upon the other available methods while the second can detect local departures from a parametric shape, as opposed to the usual goodness-of-fit tests which only track global departures. 
 A numerical study of the SCB procedures and an illustration with a speech data set are provided.  
\par

\vspace{9pt}
\par
\end{quotation}\par


\fontsize{10.95}{14pt plus.8pt minus .6pt}\selectfont
\setcounter{chapter}{1}
\setcounter{equation}{0} 
\noindent {\bf 1. Introduction}

In function estimation problems,  simultaneous confidence bands (SCB) provide a unified set of graphical and analytical tools to harness  tasks such as data exploration, model specification or validation, assessment of variability in estimation, prediction, and inference.

  In the usual setting where   the target function is observed only once at a finite number of points  with independent measurement errors, the construction of SCB has been extensively studied.  For instance, in the context of nonparametric regression on which we focus in this paper, Eubank and Speckman (1993) and Wang and Yang (2009) have used strong invariance principles to build SCB in fixed and random designs. 
Johansen and Johnstone (1990) and Sun and Loader (1994) have applied the celebrated ``tube formulas", which turn the calculation of simultaneous coverage probabilities into the simpler geometric computation of tubes' volumes, to simultaneous prediction bands and significance tests for projection pursuit regression, and to bias-corrected confidence regions with linear multivariate estimators, respectively. 
Other SCB procedures rely on bootstrapping (e.g.  Neumann and Polzehl, 1998) or on simultaneous confidence intervals (SCI) followed by interpolation arguments (e.g. Hall and Titterington, 1988). We also point to Baraud (2004) for the computation of SCI on an increasing number of (fixed) design points and to Deheuvels and Mason (2004) for SCB with asymptotic coverage level 100\%.

In the case of time series,
the construction of confidence regions proves more difficult and has received less attention in the literature. Robinson (1997) derives a SCI procedure that works under short-range, long-range and negative dependence. 
Wu and Zhao (2007) build SCB for the trend and a test for  structural breaks using a strong invariance principle. Wang (2009) provides SCB based on constant or linear splines. The case of a random design is studied in Zhao and Wu (2008).

We turn to the case of {\it functional data}, for which the statistical objects under study are viewed as functions rather than scalars/vectors and are generally observed on a dense grid. We note that this setting has attracted considerable interest over the recent years due to the now 
routine collection of high frequency data allowed by technology. 
(Numerous examples of areas of applications  can be found in 
the books of Ramsay and Silverman, 2005, and Ferraty and Vieu, 2006.) 
In the functional data framework where both the numbers of sampled functions, say $n$, and of design points, say $p$, may vary, Degras (2008) showed that all linear smoothers have an asymptotic variance of order $n^{-1}$. In the same paper SCI are derived and compared to Bonferroni- and Scheff\'e-type intervals. 
Degras (2009) builds SCB  for the regression function by coupling a functional central limit theorem [CLT] with a limit result on the supremum of a Gaussian process.
 
The present work  provides a functional asymptotic normality result that serves as a building block for estimation and  inference in nonparametric regression with functional data.
 We present two applications of this result to the band estimation of the regression function and to a goodness-of-fit test for curvilinear regression models. To the best of our knowledge, these tasks have not yet been addressed in the functional data setup. 
The proposed band estimation procedure corrects some shortcomings of Degras (2009) by fully accounting for the covariance structure of the data-generating process. 
The goodness-of-fit test relies on a SCB for the difference between the regression function and its projection onto the null space. It can detect local departures, as opposed to other tests based on residual sums of squares or $L_2$ norms that only track global departures. 

The remainder of the paper is organized as follows. 
Section 2 presents the regression model and estimator under study. 
Section 3 establishes the main result of functional asymptotic normality. 
 The normal SCB for the regression function and the goodness-of-fit test are constructed in Section 4 
 and studied numerically in Section 5 along with bootstrap SCB and a pseudo-likelihood ratio test. 
 Section 6 illustrates the use  of SCB methods with a speech data set.  
 A discussion is provided in Section 7. The proofs are deferred to the Appendix.\\
 
\par

\setcounter{chapter}{2}
\setcounter{equation}{0} 
\noindent {\bf 2. Model and local linear estimator}

Let $(Y_{ij},x_j),\, 1\leq i \leq n,\, 1\le j \le p,$ be  repeated measurements 
on a random sample of $n$ experimental  units, where  $Y_{ij}$ stands for the observed 
(scalar) response on the $i$th unit at the fixed value $x_j $ of a variable $x $. 
It is assumed that $x$ varies in  a compact subset of $\mathbb{R}^d$ for some $d\ge 1$, say $ [0,1]^d$ without loss of generality. 
Consider the regression model 
\begin{equation}\label{data model}
Y_{ij} =   \mu(x_{j}) + Z_i(x_{j}) + \varepsilon_{ij}  
\end{equation}
where $\mu$ is an unknown smooth function, 
the $Z_i$ are independent copies of a random process $Z=\{Z(x):x\in [0,1]^d \}$ with mean zero and covariance function $R$, and  the   $\varepsilon_{ij}$ are random  errors having mean zero. 
 A triangular array structure is assumed for the data as $n$ varies (in particular  $p=p(n)$ and $x_j=x_{j}(n,p)$).
The regression function $\mu$ may be viewed as a population mean response 
while the $Z_i$ represent individual departures from $\mu$.  
In this  paper we restrict our attention to the case $d\in \{Ê1,2 \} $ for simplicity but our results 
extend to higher dimensions.  The following assumptions are needed for our asymptotic study:

 \begin{itemize}
\item[(A.1)] The function $\mu$ has bounded (partial) derivatives on $[0,1]^d$ up to order 2. \vspace*{-.5mm}
\item[(A.2)]  With probability one, 
$|Z(x)-Z(x')| \leq M \| x-x' \|^\beta $ for all $x,x' \in [0,1]^d$, where $M$ is a random variable (r.v.) of finite variance, $\beta> 0$ is a constant, and $\| \cdot \|$ is a norm on $[0,1]^d$.  \vspace*{-.5mm}
\item [(A.3)]  The $x_j$ form a regular grid generated by a product density $f(t_1,\ldots,t_d)=\prod_{k=1}^d f_k(t_k)$,
where the $f_k$ are continuous and positive densities on $[0,1]$.  
It holds that 
$\{ x_j:1\le j \le p\} = \{ (x_{j_1 ,1},\ldots, x_{ j_d, d}):    1\le  j_k \le  p_k, 1\le k \le d \} $
 where  $\int_0^{x_{j_k,k}}f_k(t)dt =  \frac{ j_k- 0.5}{p_k}$. 
 In particular $p=\prod_{k=1}^d p_k$. 
 \vspace*{-.5mm}
\item[(A.4)]  
$n=o\big(\min_{k=1,\ldots,d}(p_k^4)  \big)$ 
and $n^{1/(4d)}\log(p)=o(p)$ as $n,p\to \infty$.   \vspace*{-1mm}
 \item[(A.5)]  The random vectors  $(\varepsilon_{i1},\ldots,\varepsilon_{ip})^{\top} , i=1,\ldots,n$ 
are  mutually independent, independent of the $Z_i$, and   
have the same normal distribution $N_p(0,\mathbf{V})$. The eigenvalues of the covariance matrix
$\mathbf{V}$ are uniformly bounded in  $n,p$.  
\end{itemize}
 Note that assumptions (A.1)--(A.5) are tailored for the functional data framework wherein typically, the design points are balanced and taken on a regular grid, the observed random processes are smooth, and the design size $p$ is large enough relative to the sample size $n$ so as to accommodate (A.4).

We recall here the definition of the local linear estimator (e.g.  Fan, 1992). 
Let us denote by $\langle \cdot, \cdot \rangle_d $  the euclidean scalar product 
in $\mathbb{R}^d$  and use arithmetic operations in a componentwise sense.  
Let $K$ be a kernel function on $\mathbb{R}^d$ which we take nonnegative, Lipschitz-continuous,  
with support $[-1,1]^d$ and such that $K(0)>0$. 
Let $h=(h_1,\ldots,h_d)>0$ be a vector of bandwidths. Write 
the data averages as $\overline{Y_j}=n^{-1}\sum_{i=1}^n  Y_{ij},\, 1\le j \le p$. 
For a 
given location $x\in [0,1]^d$, the local linear estimator $\widehat{\mu} (x)$ is defined as $\widehat{\beta}_0$, where  $( \widehat{\beta}_0,\widehat{\beta}_1)$ is the solution of the minimization problem
\begin{equation}\label{objective}
\min _{(\beta_0,\beta_1)\in \mathbb{R}^{d+1}}
	 \sum_{j=1}^p \Big( \overline{Y_j} - \beta_0 - \langle \, \beta_1, (x_j-x)\, \rangle_d \Big)^2 \: 
 K\left( \frac{ x_j-x}{  h} \right).
\end{equation}
This estimator can be expressed as 
\begin{equation}\label{ll_est}
\widehat{\mu} (x)   =  \sum_{j=1}^p W_j(x) \, \overline{Y_j} , 
\end{equation}
where  $W_j(x) = \displaystyle \frac{w_j(x)}{  \sum_{j=1}^p  w_j(x)}$ and 
\begin{equation}\label{poids LL dim1}
\left\{ 
\begin{array}{lcl}
w_j(x)&=&\displaystyle \frac{1}{ph}\,   \Big( s_{2}(x) - (x_j-x) s_{1}(x) \Big)\: K \left( \frac{x_j-x }{h}\right) \\
s_{l}(x)&=& \displaystyle\frac{1}{ph} \sum_{j=1}^p  (x_j-x)^l \, K \left( \frac{x_j-x }{h}\right), \quad l=0,1,2
\end{array}
 \right. \quad \textrm{when} \: d=1,
\end{equation}
or 
\begin{equation}\label{poids LL dim2} 
\left\{ 
\begin{array}{lcl}
w_j(x) \hspace*{-2mm}&=&\hspace*{-2mm}\bigg[ s_{11}(x)s_{22}(x)- s_{12}^2(x)  + \big(s_{02}(x)s_{12}(x)- s_{01}(x)s_{22}(x) \big)(x_j-x)^{(1)} \\ 
&& \displaystyle +  \big(s_{01}(x)s_{12}(x)- s_{02}(x)s_{11}(x) \big)(x_j-x)^{(2)}\bigg]  \: \frac{1}{ph_1h_2}\:K \left( \frac{x_j-x}{h}\right) \\
s_{kl}(x) \hspace*{-2mm}&=& \hspace*{-2mm} \displaystyle \frac{1}{ph_1h_2} \sum_{j=1}^p (x_j-x)^{(k)} (x_j-x)^{(l)}   K \left( \frac{x_j-x}{h}\right) ,  \quad k,l=0,1,2
\end{array}
 \right.
\end{equation}
when $d=2$, with  the notations $z=(z^{(1)},z^{(2)})$ and $z^{(0)}=1$ for all $z\in \mathbb{R}^2$. \\


\par

\setcounter{chapter}{3}
\setcounter{equation}{0} 
\noindent {\bf 3. Functional asymptotic normality}

We start with a definition. 
Let $\|\cdot \|_\infty $ be the supremum norm on the space of continuous functions $ C([0,1]^d)$. 
A sequence $(X_n)$ of random elements of $C([0,1]^d)$ is said to converge weakly to a limit $X$ in  $ C([0,1]^d) $ 
if $\mathbb{E}(\phi(X_n)) \to \mathbb{E}(\phi (X))$ as $n\to\infty$  for all bounded, uniformly continuous functional $\phi$ on $ (C([0,1]^d),\|\cdot \|_\infty )$ (e.g. Pollard, 1990, p.44). 
Let $\mathcal{G}(m,C)$ denote a real-valued Gaussian process indexed by $[0,1]^d$ with  arbitrary mean and covariance functions $m$ and $C$. 
We are now in position to state the main result.

\begin{theorem}\label{main result} 
Assume (A.1)--(A.5) in model \eqref{data model} with $d\in \{1,2\}$. 
Consider the  local linear estimator $\widehat{\mu} $ defined in \eqref{ll_est}--\eqref{poids LL dim2} with a bandwidth $h=h(n,p)$ such that:  $n\|h\|^4\to 0$  and $ (p/\log(p))   \prod_{k=1}^d  h_k \to \infty$ 
as $n,p \to\infty$. 
Then 
$\sqrt{n} (\widehat{\mu} - \mu) $ converges weakly to $ \mathcal{G}(0,R)$ in $C([0,1]^d)$. 
 \end{theorem}

\noindent {\it Remarks.}
\begin{enumerate}

\item {\it Convergence rate.} 
The 
fact that the normalizing rate $\sqrt{n}$ depends neither  on  $p$ nor on $h$ 
is consistent with the literature. It reflects the fact that $(\widehat{\mu}-\mu)$ is essentially a smoothed version of $\overline{Z}=n^{-1}\sum_{i=1}^n Z_i$,  whose covariance structure ($R/n$)  is essentially unaffected by smoothing or discretization. 
 
\item {\it Regularity of $Z$.} The conclusion of Theorem 1 holds under weaker conditions than the stochastic H\"older continuity
(A.2), 
e.g. when $Z$ is mean-square continuous and has bounded variations (Degras, 2009). 
However (A.2) is needed for the SCB and test procedure of Section 2. 


\item {\it Joint growth of $n$ and $p$.} The growth conditions (A.4) enforce that (a power of) $p$ is large enough relative to $n$. 
This ensures the existence of a bandwidth $h=h(n,p)$ small enough to make the squared bias of $\widehat{\mu}$ 
  negligible before its variance but large enough to smooth out 
   the measurement errors $\varepsilon_{ij}$ uniformly  as $n,p\to \infty$. 
(A.4) can be weakened (allowing for larger $n$ and smaller $p$) by assuming a higher order of differentiability for $\mu$ in (A.1) and using higher order local polynomial estimators or bias reduction techniques.

  \item {\it Measurement errors.} The uniform bound on the covariance matrix $\mathbf{V}$
 in (A.5) accommodates various forms of dependence such as short-range dependence and ARMA or mixing processes. The normality assumption is not essential to the results, however the decrease rates in the tail probabilities of the $\varepsilon_{ij}$  influence the size of $h$ needed to smooth out these errors. 
 For instance, if the normality assumption is dropped then the factor $(p/\log(p))$ in the condition $ (p/\log(p))   \prod_{k=1}^d  h_k \to \infty$ in Theorem 1 becomes $ p^{1/2} $ whereas if the $\varepsilon_{ij}$ are assumed to be uniformly bounded, the factor  is equal to  $p$. 
  
 \item{\it Bandwidth selection.} For simplicity Theorem 1 is presented with a deterministic $h$ but it also holds when $h$ depends on the data $(Y_{ij},x_j)$ in a way that $ C_1 a(n,p) \le h(n,p) \le C_2 a(n,p)$, where $a(n,p)$ is a deterministic sequence satisfying the conditions of Theorem 1 and $C_1,C_2>0$ are constants. Hence, suitable plug-in or cross-validation methods  (e.g. Hart and Wehrly, 1993) can be used to select
 $h$. Also note that in the present context of functional data, the asymptotic variance of $\widehat{\mu}(x)$ is of order $n^{-1}$ and is only affected by 
 $h$ through a second-order term in $\mathcal{O}(hn^{-1})$ (see {\it ibid.}).  
 Therefore, the strategy adopted here to render the bias  (of order $h^4$) negligible before the variance 
 is compatible with the optimization in $h$ of the asymptotic mean squared error of $\widehat{\mu} (x)$: 
  it does not slow down the convergence.

    \item {\it Longitudinal data.} In contradistinction to the functional data setup, asymptotic normality results in $C([0,1]^d)$ cannot be obtained in the longitudinal data setup where typically, many random functions are observed each at a few time points. 
In this setup, nonparametric estimators mostly average data across the sample units, which are independent, and not within,  where the random process structure plays. 
 Therefore they converge pointwise to a Gaussian white noise process at the usual regression rates (Yao, 2007).     

\end{enumerate}

%

\bigskip 
\par

\setcounter{chapter}{4}
\setcounter{equation}{0} 
\noindent {\bf 4. Applications}

\par
\noindent {\bf 4.1 Simultaneous confidence bands for $\mu$}

We apply here Theorem \ref{main result} to the construction of SCB for the regression function $\mu$. 
Let us denote respectively by $\sigma^2$ and $\rho$ the variance and correlation functions of $Z$. 
Without loss of generality, we assume that $\sigma^2$ is positive over $[0,1]^d$ so that $\rho$ is well-defined. 
It stems from Theorem \ref{main result} and Slutsky's theorem that for any uniformly consistent estimator $\widehat{\sigma}^2$ of $\sigma^2$, the standardized estimator $\sqrt{n} \big( \widehat{\mu} - \mu \big)/\widehat{\sigma}$ converges to  $\mathcal{G}(0,\rho)$ in $C([0,1]^d)$ as $n\to\infty$. For a given confidence level $1-\gamma$,  
we seek approximate SCB of the form 
 \begin{equation}\label{SCB}
  \left[ \widehat{\mu}(x) - c_\gamma \, \frac{\widehat{\sigma}(x)}{ \sqrt{n}}\,,\, \widehat{\mu}(x) + c_\gamma \, \frac{\widehat{\sigma}(x)}{ \sqrt{n}} \right] , \quad x \in [0,1]^d ,
\end{equation}
where  $\mathbb{P}\big( \| \mathcal{G}(0,\rho) \|_\infty > c_\gamma \big) \approx \gamma $.

 A convenient estimator of $\sigma^2$ is the empirical variance function of the smooth curves $\widehat{\mu}_i = \sum_{j=1}^p W_j Y_{ij} , \,i=1,\ldots,n ,$ namely 
\begin{equation}\label{empirical variance}
\widehat{\sigma}^2(x) =  \frac{1}{n-1} \sum_{i=1}^n \big( \widehat{\mu}_i (x) - \widehat{\mu} (x) \big)^2.   
\end{equation}
This  estimator is unbiased for the finite sample variance $n \var(\widehat{\mu}(x))$ and  it converges uniformly to $\sigma^2(x)$ in probability. 
(The uniform convergence is obtained  by exploiting (A.2)-(A.5) together with a uniform law of large numbers (e.g. Pollard, 1990, Th. 8.2) 
and a classical limit result on the largest eigenvalue of a Wishart matrice (Geman, 1980) to control uniformly the errors $\varepsilon_{ij}$.)

Two difficulties arise in the computation of the threshold $ c_\gamma$: first, the correlation function $\rho$  must be estimated and second, 
even if $\rho$ were known, there exists no formula for the distribution of the maximum of a general Gaussian process (see e.g. Adler, 1990, p.5). 
  For the first problem a suitable estimator of $\rho$  is  the empirical correlation function 
  \begin{equation}\label{empirical correlation}
\widehat{\rho}(x,x') =
 \frac{ \sum_{i=1}^n  \widehat{\mu}_i (x)\, \widehat{\mu}_i (x') - n\,  \widehat{\mu} (x)\,  \widehat{\mu}  (x') }
 {(n-1) \, \widehat{\sigma}(x) \,  \widehat{\sigma}(x') } \,. 
\end{equation}
For the second problem we  resort to  numerical techniques to estimate $c_\gamma$. 
(See Section 6 for a discussion of the limitations of theoretical approximations to the distributions of maxima of Gaussian processes.) 
This can be done by simulating, conditional on $\widehat{\rho}$, 
a large number of sample paths of $\mathcal{G}(0,\widehat{\rho})$ in order to obtain the law $\mathcal{L}( \| \mathcal{G}(0,\widehat{\rho}) \|_\infty | \widehat{\rho})$ and then by setting  $c_\gamma$ as the associated $(1-\gamma)100\%$ quantile: 
\begin{equation}\label{threshold}
\mathbb{P}\big( \| \mathcal{G}(0,\widehat{\rho}) \|_\infty > c_\gamma | \widehat{\rho} \big) = \gamma .
\end{equation}
The fact that $c_\gamma$ satisfies approximately $\mathbb{P}\big( \| \mathcal{G}(0,\rho) \|_\infty > c_\gamma \big) = \gamma $ is  
justified by the convergence of  $ \mathcal{G}(0,\widehat{\rho})$ to $\mathcal{G}(0,\rho)$ 
in $C([0,1]^d)$, conditionally on $ \widehat{\rho}$. This in turn  stems from: (i)  
 the finite-dimensional convergence  of  $ \mathcal{G}(0,\widehat{\rho})$ thanks to the uniform convergence of $\widehat{\rho}$, and (ii) the  
  asymptotic tightness of  $ \mathcal{G}(0,\widehat{\rho})$ obtained through entropy calculations very similar to those in the Appendix.   

Gathering the previous elements, we obtain the following result.

\begin{theorem}\label{SCB theorem}
Under the assumptions of Theorem \ref{main result}, the simultaneous confidence bands  \eqref{SCB} 
have asymptotic coverage level $1-\gamma$  for $\mu$: 
\[ \lim_{n,p\to\infty} \mathbb{P} \left(   \widehat{\mu}(x) - c_\gamma \, \frac{\widehat{\sigma}(x)}{ \sqrt{n}} \le \mu(x) \le  \widehat{\mu}(x) + c_\gamma \, \frac{\widehat{\sigma}(x)}{ \sqrt{n}} , \quad x \in [0,1]^d \right) = 1-\gamma \]
 with the estimators  $\widehat{\sigma}$, $\widehat{\rho}$, and the threshold $c_\gamma$ 
 defined in \eqref{empirical variance}-\eqref{empirical correlation}-\eqref{threshold}.
\end{theorem}

 
In the case where the sample size $n$ is small and the process $Z$ cannot be assumed to have an approximate normal distribution, it may not be reasonable to rely on a functional CLT  
to build SCB for $\mu$. We thus propose, without theoretical justification, the following naive bootstrap procedure: 
\begin{enumerate}
\item Resample with replacement from the $ \widehat{\mu}_i , i=1,\ldots,n $ to produce a bootstrap sample $\mu^\ast_1 ,\ldots, \mu^\ast_n $. 
\item Compute the empirical mean and variance functions of the $\mu^\ast_i$, 
say $\mu^\ast $ and $(\sigma^\ast)^2$, and compute  $z^\ast=\sqrt{n} \| (\mu^\ast - \widehat{\mu})  / \sigma^\ast \|_\infty $. 
\item Repeat steps 1 and 2 many  times to approximate the conditional law 
$\mathcal{L}^\ast = \mathcal{L}(z^\ast | Y_{ij}\textrm{'s} )$ 
 and take the $(1-\gamma)100\%$ quantile of $\mathcal{L}^\ast$ for 
 $c_\gamma$ in   \eqref{SCB}. 
\end{enumerate}


\medskip

\par
\noindent {\bf 4.2 A goodness-of-fit test for parametric models}\label{goodnessfit}

We now apply the ideas underlying Theorems \ref{main result}-\ref{SCB theorem} 
to a goodness-of-fit test for curvilinear regression models.  
Indeed, with the knowledge of the limit distribution of an estimator 
 in  $ (C([0,1]^d),\|\cdot \|_\infty )$, 
 it becomes possible 
to detect and test local departures from a given candidate model for $\mu$.  
 This feature should be contrasted with tests  based on euclidean norms which only track global departures. 
See e.g. Azzalini and Bowman (1993), H\"ardle and Mammen (1993), and Stute (1997) where parametric and nonparametric estimates are compared either via their residual sum of squares or directly through $L_2$ distances.

 Consider a candidate parametric model for $\mu$ of the form  
\begin{equation}\label{param model}
\quad \mu  \in  \mathcal{M} =\left\{   \sum_{l=1}^{L} \theta_l \varphi_l : (\theta_1,\ldots, \theta_L)\in \Theta \right\} 
\end{equation}
 where $L\leq 1$ is a fixed integer,  
$\Theta\subset \mathbb{R}^L  $ is a parameter space, and 
 $(\varphi_1,\ldots, \varphi_L)$ is a family of functions on $[0,1]^d$ 
 satisfying
 \begin{enumerate}
\item[(B.1)] The   $\varphi_l$ are  orthogonal w.r.t.  
 the inner product $ \langle  g_1,g_2 \rangle_f = \int g_1(x) g_2(x) f(x) dx$.
\item[(B.2)] The $\varphi_l$ have bounded (partial) derivatives on $[0,1]^d$ up to order $2$. 
\end{enumerate}
Introducing the vectors  $\overline{Y}=(\overline{Y_1},\ldots,\overline{Y_p})^{\top}$, 
$\varphi(x) =(\varphi_1(x),\ldots,\varphi_L(x))^{\top}$ and the $p\times L$  matrix
$\mathbf{\Phi}= ( \mathbf{\varphi}(x_1),\ldots, \mathbf{\varphi}(x_p) )^{\top}$, 
the least squares estimator of $\mu(x)$ under \eqref{param model} reads

\begin{equation}\label{OLS fit}
\widehat{\mu}_{LS}(x) =  \varphi(x)^{\top} (\mathbf{\Phi^{\top} \Phi})^{-1} \mathbf{\Phi^{\top}} \overline{Y}.
\end{equation}
\vskip 2mm

We now apply the local linear weights $W(x)$\,$=$\,$(W_1(x),\ldots,W_p(x))^{\top}$ 
to the residuals of the parametric fit \eqref{OLS fit}. 
The smoothed residual random process $r$ is 
\begin{equation}\label{residual process}
r(x) = \sum_{j=1}^p W_j(x)   \big(\overline{Y}_j - \widehat{\mu}_{LS}(x_j) \big)  = 
W(x)^{\top} \left( \mathbf{I -P} \right) \overline{Y}
\end{equation}
where $\mathbf{I}$ denotes the $p\times p$ identity matrix and $\mathbf{P =\Phi(\Phi^{\top}\Phi)^{-1}\Phi^{\top} }$ denotes the $p\times p$ projection matrix onto the space spanned by the columns of $\mathbf{\Phi}$.

Next, we determine the asymptotic mean and covariance functions of the (scaled) process $r$. 
Under  \eqref{param model}, it is straightforward to see that $\mathbb{E}(r(x))=0$. 
More generally let $P$ be the orthogonal projection from  $(L_2([0,1]^d),\langle \cdot, \cdot \rangle_f )$ onto the linear subspace $\mathcal{M}$ and by $\theta=(\theta_1,\ldots,\theta_L)^{\top}=(\langle \varphi_1, \mu\rangle, \ldots, \langle \varphi_L, \mu\rangle )^{\top}$ the vector of coefficients of  $P\mu$ in $\mathcal{M}$.  
Observe that 
$ \| \mathbb{E} (\widehat{\mu}) -  \mu \|_\infty = \| \mu'' \|_\infty \, \mathcal{O}(\| h \|^2 ) $ 
by (A.1) and the bias properties of local linear estimators. 
Also, exploiting the former bias properties, (B.1),  (B.2), and classical error bounds for numerical approximations of  integrals, it can be easily proved that 
\begin{align}
\mathbb{E}\big(W(x)^{\top} \mathbf{P} \overline{Y} \big)  & = (W(x)^{\top} \mathbf{\Phi) (\Phi^{\top}\Phi)^{-1}(\Phi^{\top} }\mathbb{E} (\overline{Y}) )\nonumber \\
& =\big[ 1+  \| \mu'' \|_\infty \mathcal{O}( \| h \|^2) \big] \, \varphi(x)^{\top} \: \big[\mathbf{I} + \mathcal{O}( p^{-1}) \big] \: \theta \big[1 + \mathcal{O}( p^{-1}) \big] \nonumber \\
& = \varphi(x)^{\top}  \theta +   \mathcal{O}(\| \mu'' \|_\infty \| h \|^2+ p^{-1}) \nonumber \\ 
& = P\mu(x)  +   \mathcal{O}(\| h \|^2 \| \mu'' \|_\infty + p^{-1})  \nonumber
\end{align}
 uniformly in $x\in [0,1]^d$. Combining these relations with \eqref{residual process} yields
\begin{equation}\label{bias smooth res proc}
\mathbb{E}(r(x)) = \mu(x) - P\mu(x) + \mathcal{O}(\| \mu'' \|_\infty \| h \|^2 + p^{-1} ).
\end{equation}

\noindent With the above calculations one can infer from \eqref{residual process} that $r$ has the same asymptotic covariance as the process $\widehat{\mu} - \widehat{\mu}_{LS}$ and then, 
using these calculations together with Theorem \ref{main result},  
 the limit covariances and cross-covariances of $\widehat{\mu} $ and $\widehat{\mu}_{LS}$ (scaled by $\sqrt{n}$) 
are derived without difficulty. (In particular 
the limit covariance of $ \sqrt{n}\widehat{\mu}$ is $R$.) 
Finally  the limit covariance function of $\sqrt{n} r$ is 
\begin{equation}\label{limit covariance for smoothed residual}
\begin{split}
 \hspace*{-3mm}\Gamma (x,x')=R(x,x')+     \sum_{ k=1}^ L \sum_{l=1}^L \varphi_k(x) \varphi_l(x') \iint R(u,v) \varphi_k(u)\varphi_l(v) f(u)f(v) du dv   \\
 -  \sum_{l=1}^L \Big(  \int   R(x,u)\varphi_l(u)f(u)du \Big) \, \varphi_l(x')-  \sum_{l=1}^L \Big(  \int   R(x',u)\varphi_l(u)f(u)du \Big) \, \varphi_l(x)
\end{split}
\end{equation}
where the simple (resp. double) integrals are taken over $[0,1]^d$ (resp. $[0,1]^{2d}$). 

Denote by $\mathcal{M}^c$ the orthogonal complement of $\mathcal{M}$ in $(L_2([0,1]^d, \langle \cdot,\cdot \rangle_f)$ and by $C^2([0,1]^d)$ the space of functions having continuous partial derivatives on $[0,1]^d$ up to order $2$. 
Using the proof techniques of Theorem \ref{main result}, 
the following asymptotic normality holds true for $\sqrt{n}r$. 

\begin{theorem}\label{as-norm for test stat}
Assume (A.1)--(A.5) and (B.1)-(B.2) in model \eqref{data model} with $d$\,$\in$\,$\{1,2\}$.  
Consider the smooth residual process $r$ defined in \eqref{residual process} 
with a bandwidth $h$ satisfying:  $n\|h\|^4 \to 0$ and $ (p/\log(p))   \prod_{k=1}^d  h_k \to \infty$
as $n,p \to\infty$. \\
Then under the null hypothesis \eqref{param model}, 
 $\sqrt{n} r$ converges weakly to 
 $\mathcal{G}(0,\Gamma)$ in $C([0,1]^d)$. 
 Under the sequence of local alternatives
 $ \mu= \varphi^{\top}\theta +  g/\sqrt{n}$, where  $\theta \in \Theta$ and $g\in C^2([0,1]^d)\bigcap  \mathcal{M}^c$ are fixed,  
$\sqrt{n} r $ converges weakly to  $\mathcal{G}(g,\Gamma)$ in $C([0,1]^d)$. 
\end{theorem}

We now apply Theorem \ref{as-norm for test stat} to  testing \eqref{param model} against fixed or local alternatives in a way 
that strictly parallels the SCB construction of 
Section 4.1. In particular it will be necessary to estimate the variance and correlation functions 
$\sigma_\Gamma$ and $\rho_\Gamma$ of $\sqrt{n} r $ as well as a threshold $c_\alpha$ for a related Gaussian process. 
First note that for finite samples, the covariance function of $\sqrt{n} r $ is 
\begin{equation}\label{cov r finite sample}
\Gamma_n(x,x') =  W(x)^{\top}  \left(\mathbf{I - P } 
\right)\left( \boldsymbol{\Sigma } 
+ \mathbf{V}   \right)  \left(\mathbf{I - P  }\right)    W(x')^{\top}
\end{equation}
where $\boldsymbol{\Sigma } $ is the $p\times p$ covariance matrix $(R(x_j,x_k))$ and $\mathbf{V}$ is the common 
covariance matrix of the measurement errors in (A.5). The matrix $\left( \boldsymbol{\Sigma } + \mathbf{V}   \right)$
 can be estimated by the empirical covariance of the data $Y_{ij}$, which is then plugged in  \eqref{cov r finite sample} to produce an 
   estimator $ \widehat{\Gamma}(x,x')$ of $ \Gamma(x,x') $. The related variance and correlation estimators are 
    $ \widehat{\sigma}_\Gamma(x)=  \widehat{\Gamma}(x,x)^{1/2}$ and 
   $\widehat{\rho}_\Gamma(x,x') =  \widehat{\Gamma}(x,x')/ \big( \widehat{\sigma}_\Gamma(x)  \widehat{\sigma}_\Gamma(x') \big)$. 
Now for a given significance level $\alpha$, a threshold $c_ \alpha $ such that $ \mathbb{P}(\|\mathcal{G}(0,\rho_\Gamma) \|_\infty > c_ \alpha \big) \approx  \alpha $ may be found exactly as in Section 4.1 by numerical simulation of 
$  \|\mathcal{G}(0, \widehat{\rho}_\Gamma)\|_\infty$ conditional on $\widehat{\rho}_\Gamma$ followed by the computation 
of the $(1-\alpha)100\%$ quantile of the resulting distribution: $ \mathbb{P}(\|\mathcal{G}(0,\widehat{\rho}_\Gamma) \|_\infty > c_ \alpha \big| \widehat{\rho}_\Gamma \big) = \alpha $.

From \eqref{bias smooth res proc} and Theorem \ref{as-norm for test stat}, we  deduce the following result.
\begin{corollary}\label{test consistency}
In model \eqref{data model}, consider the candidate model  \eqref{param model} for $\mu$ and the test statistic $T =\sqrt{n} \big\| r/\widehat{\sigma}_\Gamma \big\|_\infty $ defined by 
\eqref{residual process}-\eqref{cov r finite sample}. 
 For a given $\alpha\in (0,1)$, let $c_\alpha$ be the conditional quantile defined above.\\
Under the assumptions of Theorem \ref{as-norm for test stat}, 
the test obtained by rejecting \eqref{param model} if  $T >c_\alpha $ has asymptotic significance level $\alpha$ and 
is consistent against any fixed alternative $H_1: \mu=g  \in C^2([0,1]^d)\bigcap  \mathcal{M}^c$. 
Given a constant $ B>0$ and a real sequence $\epsilon_n> 0$ such that 
  $n^{-1/2} = o(\epsilon_n)$, the test is also consistent against the sequence of local alternatives 
$H_n:  \mu \in \big\{ g\in C^2([0,1]^d): \| g'' \|_\infty \leq B, \| g - Pg \|_\infty = \epsilon_n \big\}$. 
\end{corollary}
\noindent Note that graphically, the test can be interpreted as plotting the SCB $\big[ r(x) \pm  c_\alpha \, \widehat{\sigma}_\Gamma (x)/\sqrt{n}) \big]$ for $\mu(x) - P\mu(x)$ and rejecting \eqref{param model} if the horizontal line $y=0$ is not contained within the bands. \\


\par

\setcounter{chapter}{5}
\setcounter{equation}{0} 
\noindent {\bf 5. Numerical study}

\par
\noindent {\bf 5.1 Normal and bootstrap SCB procedures}

In this section we assess the normal and bootstrap SCB procedures of Section 4.1 in terms of coverage and  amplitude through the numerical study of two examples of model \eqref{data model}. 
In short, the first example depicts a favorable situation with a smooth polynomial trend, Gaussian data, and no measurement errors while the second features very adverse conditions with a rapidly varying trend, a strongly non-normal random process $Z$, and additive white noise. 
More specifically, the first example, taken from Hart and Wehrly (1986), is  
\begin{equation}\label{model1}
\left\{ \begin{array}{l}
Y_{ij} =  \mu(x_j) + Z_i(x_j)  , \quad 1\leq i \leq n , 1 \leq j \leq p, \\
\mu(x)= 10 x^3 - 15 x^4 + 6x^5,\\
x_j=(j-0.5)/p, \\
Z_i  \stackrel{iid}{\sim} \mathcal{G}(0,R) \: \textrm{with } R(x,x') = (0.25)^2 \exp( 20 \log(0.9) | x-x'|).
\end{array} \right.
\end{equation}


\noindent 
Here, the $x_j$ are equidistant and the $Z_i$ are distributed as a centered Gaussian process with an Ornstein-Uhlenbeck covariance function chosen so that any two measurements spaced by 0.05 units have correlation 0.9. The noise level $\sigma=0.25$ represents 25\% of the range of the trend $\mu$, which is considered as a moderate amount of noise in the data. 
The second model is specified by 
\begin{equation}\label{model2}
\left\{ 
\begin{array}{l}
Y_{ij} = \mu(x_j) + Z_i(x_j) +  \varepsilon_{ij} , \quad 1\leq i \leq n , 1 \leq j \leq p, \\
\mu(x)=\sin(8\pi x) \exp(-3 x) , \:
x_j=(j-0.5)/p, \\
Z_i  \stackrel{iid}{\sim}  Z \: \textrm{with } Z(x) = (\sqrt{2}/6) \left( \eta_1-1\right) \sin(\pi x) +(2/3) \left(\eta_2-1 \right)( x - 0.5), \\
\eta_1 \sim \chi^2_1, \:\textrm{and } \eta_2 \sim \textrm{Exponential} (1), \\
\varepsilon_{ij}  \stackrel{iid}{\sim} N(0,0.1^2) , \quad \varepsilon_{ij} \textrm{ and } Z_i \textrm{ independent}.
\end{array}
\right.
\end{equation}

\noindent In this case, the regression function $\mu$ displays rapid variations over $[0,1]$ and has a sharp peak near the origin at $x=0.058$. The process $Z$ strongly deviates from normality, being based on chi-square and exponential r.v.. 
 The standard deviation function $\widetilde{\sigma}(x) = (R(x,x) + 0.1^2 )^{1/2}$ ranges between  0.295 and 0.348, which represents a fraction between 21\% and 25\% of the range of $\mu$. 
 However, looking at the local variations of $\mu$ as measured by $| \mu''|$ and the noise level $\widetilde{\sigma}$, it appears that $ | \mu''| / \widetilde{\sigma} $ ranges in $[7, 1650]$  (compare with the range $[ 0,23 ]$ for the same function ratio in \eqref{model1}). Such a range indicates than in regions where $\mu$ has high curvature, i.e. around peaks and troughs, 
 serious estimation problems should arise due to the fact that the (squared) bias will be overwhelmingly larger than the variance, a violation of the conditions of Theorem \ref{main result}.  In particular near $x=0.058$, the problem will be prominent since the classical peak underestimation problem will combine with  boundary effects.

The simulations were conducted in the R environment  as follows. For each model \eqref{model1} or \eqref{model2}, several values were selected for the sample size $n$, the design size $p$, and the bandwidth $h$ of the local linear estimator $\widehat{\mu}$. For each $(n,p,h)$ the model was simulated $N_{rep}=50,000$ times to assess the normal bands and only  $N_{rep}=5,000$ times for the bootstrap  due to its heavy computational cost. 
The bands were built at the confidence level $1-\gamma=95\%$ and their coverage levels (i.e. the proportion of simulations for which the bands contained  $\mu$) 
and amplitudes (in terms of the threshold $c_\gamma$ of \eqref{SCB}) 
were recorded. In model \eqref{model1} the margins of error in the coverage levels can be evaluated  as about $\sqrt{\gamma (1-\gamma)/N_{rep}}= 0.0009, \, 0.0031$ for the normal and bootstrap procedures, respectively. 
   In model \eqref{model2} the observed coverage levels are quite different from the target $95\%$ and it seems more reasonable to evaluate the margins of errors by $1/(2\sqrt{N_{rep}})=0.0022, \, 0.0070$ respectively. 
At another level, it has been observed 
that for a given setup $(n,p,h)$,  
    the main source of variability in the bands' amplitude lies in 
    the estimation of  $c_\gamma$
    whereas the estimation of $ \sigma^2(x)$ bears little influence. 
    This is why the bands' average amplitudes 
   are displayed
   in terms of $c_\gamma$, which besides allows for a direct comparison with the correct thresholds yielding nominal coverage.

The  SCB  were implemented  as follows.
For the normal SCB,  $\mu$ was estimated by a local linear fit with the Epanechnikov kernel $K(x)=0.75 \max(1- x^2,0)$.  For this task a R script based on sparse matrix representations was written by the author, allowing for fast and exact evaluations. The variance function $\sigma^2$ of $Z$ was estimated by the empirical variance function of the $\widehat{\mu}_i$ as in Section 4.1. 
The correlation function $\rho$ of $Z$ was estimated by a shrinkage estimator  $\widehat{\rho}$ based on the empirical correlation of the $\widehat{\mu}_i$  (R package \verb@corpcor@).
After that, a number $N$ of sample paths of the process $\mathcal{G}(0,\widehat{\rho})$ were simulated on an equispaced grid of size 100 in $[0,1]$ and the threshold $c_\gamma$ in \eqref{SCB}  was computed as the $95\%$-quantile of the associated sup norms ($N$ was set to $8000, 10000,$ and $13000$ for $p=10,20,100$, respectively, to ensure a good tradeoff between numerical accuracy and computational time). 
Concerning the bootstrap SCB, $\mu$ and $\sigma^2$ were estimated as in the normal SCB procedure and 
the threshold $c_\gamma$ was estimated as in Section 4.1 with 2500 bootstraps.

\begin{table}[h]
\begin{center}
\begin{tabular}{ c c c | cc| cc | c c }
&&& \multicolumn{2}{c |}{Normal SCB}  &  \multicolumn{2}{c |}{Bootstrap SCB} &  \multicolumn{2}{c}{Correct 95\% threshold $c_\gamma$} \\
\cline{4-9}
n& p & h &Coverage  & $c_\gamma$  &  Coverage & $c_\gamma$& $R$ estimated & $R$ known \\
\hline
10 & 10 & 0.2 & 0.926& 3.20 & 0.977 & 4.14 & 3.51   &2.66  \\
20 & 20 & 0.1 &  0.957 & 3.11 & 0.964 &3.19 &3.07  &2.69\\
20 & 20 & 0.15 & 0.955&3.08 & 0.961 & 3.10 &3.01 &2.61\\
20 & 20 & 0.2 &  0.951 &3.05 & 0.948 &3.03 &3.08&2.72\\
50 & 50 & 0.05 & 0.962 &3.01  & 0.947& 2.90 & 2.90  &2.73\\
50 & 100 & 0.05 &0.961 &3.00 & 0.953&2.91 &2.89 &2.69\\
50 & 100 & 0.1 & 0.970 &2.95  &0.947 &2.82 &2.82 &2.60\\
100 & 10 &0.16 &  0.931 &2.84  & 0.903&2.70 &2.98 &2.94 \\
100 & 10 &0.2 &  0.879 &2.81  &0.838&2.67& 3.23 &3.03\\
100& 20 & 0.1 &  0.960 & 2.88 & 0.966 &2.75 &2.79&2.69\\
100& 20 & 0.15 &  0.941 & 2.83 & 0.927&2.69 &2.88 & 2.72 \\
100& 20 & 0.2 &  0.892 & 2.80  &0.874 &2.65 & 3.16 &2.89\\
100& 50 & 0.05 &  0.961 &  2.93 & 0.946 &2.82 &2.82 &2.73\\
100& 100 & 0.05 &  0.961  &2.92  & 0.952 &2.82  &2.82  &2.70 \\
100 & 100 & 0.1 & 0.960  &2.87 &0.945  & 2.73 &2.77&2.62 \\
100 & 100 & 0.15 &  0.945  &2.82 &  0.946 & 2.68 &2.86& 2.65 \\
\end{tabular}
\end{center}
\caption{{\it Observed coverage levels and thresholds for SCB of nominal level $95\%$ in model \eqref{model1}. 
 For each $(n,p,h)$, the model was simulated 50,000 and 5,000 times for the normal and bootstrap procedures, respectively. The two columns $c_\gamma$ indicate the median threshold obtained in  \eqref{SCB}. 
 The last two columns show the actual thresholds yielding 95\% coverage when the covariance function $R$ is estimated or  known. }}
 \label{scb_proc_model1}
\end{table}

It can be observed from Table \ref{scb_proc_model1} that both the normal and bootstrap SCB methods work quite well in model \eqref{model1},  for a wide range of combinations of $n,p$ and $h$. They have similar performances (see Figure 1) and achieve  a coverage near the target level 95\%.  
This positive result can be explained by three favorable aspects of  \eqref{model1}: 
(i) the low curvature of the polynomial function $\mu$; 
(ii) the absence of measurement errors; and (iii) the normality of $Z$. 
The first point ensures that even for large $n$ and small $p$, the squared bias of $\widehat{\mu}$ remains uniformly small before its variance over $[0,1]$.  
The second point allows the use of small bandwidths since no smoothing is needed to control the absent errors. 
(In this case the second condition in (A.4) and the condition $ (p/\log(p))   \prod_{k=1}^d  h_k \to \infty$ in Theorem 1 can be dropped.) 
The second and third points imply, for the normal bands,  that the normal approximation to the distribution of $\widehat{\mu}$  is exact. 

\pagebreak 

 \begin{table}[htbp]
\begin{center}
\begin{tabular}{ c c c | cc| cc | c c }
&&& \multicolumn{2}{c |}{Normal SCB}  &  \multicolumn{2}{c |}{Bootstrap SCB} &  \multicolumn{2}{c}{Correct 95\% threshold $c_\gamma$} \\
\cline{4-9}
n& p & h &Coverage  & $c_\gamma$  &  Coverage & $c_\gamma$& $R$ estimated & $R$ known \\
\hline
10& 20 & 0.08 & 0.156 & 3.29 &  0.663 & 6.10  &10.31 & 5.41\\
10& 50 & 0.035 &0.769 &3.29  & 0.959 & 6.95   &6.16& 2.98 \\
10& 50 & 0.05 &0.713  & 3.28 & 0.941 & 6.50 &6.40  & 3.44 \\
15&20 & 0.08 & 0.055 &3.24 &0.341 &  4.47  &10.23& 6.22 \\
20 & 20 &0.08 & 0.013  &3.20 &  0.131  & 3.87 &10.53&6.92  \\
20 & 50 & 0.035 &0.839& 3.20 & 0.928 & 4.12  &4.48 & 3.13 \\
20& 50 & 0.05 &  0.704 & 3.19 & 0.852  & 4.02  &4.91& 3.90\\
20 & 100 & 0.02 & 0.878  & 3.20  &  0.934 & 4.22  &  4.28 &  2.78 \\
20& 100 & 0.05 &   0.716  &3.18 &  0.842 & 3.93 &4.94& 3.84 \\
50 & 50 & 0.035 & 0.814  &3.07  & 0.837 & 3.20  &3.93 & 3.59 \\
100 & 100 & 0.02 & 0.929 & 2.97  &0.927 & 2.97 & 3.18 & 2.96  \\
100 & 100 & 0.035 & 0.738 & 2.94  & 0.715 & 2.88  &3.81 & 3.82  
\end{tabular}
\end{center}
\caption{{\it Observed coverage levels and thresholds 
 for SCB of nominal level $95\%$ in model \eqref{model2}. 
 For each $(n,p,h)$, the model was simulated 50,000 and 5,000 times for the normal and bootstrap procedures, respectively. The two columns $c_\gamma$ indicate the median threshold obtained in  \eqref{SCB}. The last two columns show the actual thresholds yielding nominal coverage when the covariance $R$ is estimated or  known. }}
 \label{scb_proc_model2}
\end{table}

In model \eqref{model2} the estimation conditions are very adverse, as seen earlier. 
It is thus no surprise to observe in Table \ref{scb_proc_model2} that the coverage levels fall short of the 95\% target level 
both for normal and bootstrap SCB, although the bootstrap is more robust. On the other hand the last two columns of Table 2 show how intrinsically difficult the band estimation is in \eqref{model2}. For instance when $p=20$ and $R$ is unknown, the threshold yielding correct coverage is close to 10 (compare to the 95\% standard normal quantile 1.96 used in  pointwise confidence bands), yielding SCB so large that they loose all practical interest. (See also the right panel in Figure 1.)
Note that the extreme difficulty of the case $p=20$ stems mostly from the sparsity of the data near the sharp peak of $\mu$ at $x=0.058$.        
Regarding the influence of smoothing on the coverage level, it appears in Table \ref{scb_proc_model2} that the smaller the bandwidth $h$, the higher the coverage. This observation is essentially related to the control of the bias and it has been confirmed with a wider range of values $h$ not displayed here. For each $p=20,50,100,$ the values selected for $h$ were first, the smallest $h$ for which $\widehat{\mu}$ is well-defined on the evaluation grid and second, a nearby value indicating how quickly the coverage degrades when $h$ increases.   
Interestingly enough, increasing the sample size $n$ has different effects on the coverage according to $p$: 
for $p=20,$  as $n$ increases the coverage decreases. This is due to the corresponding decrease in $ \var(\widehat{\mu}(x)) \approx \sigma^2(x)/n$, which makes the squared bias increasingly non-negligible before the variance. For $p=50$, increasing $n$ also increases the squared bias to variance ratio but the latter may remain negligible provided that $n$ is not too big: the coverage increases from $n=10$ to $n=20$ and then decreases from $n=20$ to $n=50$. For $p=100$ the coverage, as a function of $n$, would start to decrease after  
a value $n$ much larger than 100. (Note that the supremum of the squared bias to variance ratio is asymptotic
to $nh^4/4 \cdot \| \mu'' / \sigma^2 \|_\infty$.) 
Two other important effects of increasing $n$ are first to reduce the stochastic error in the normal approximation to the distribution of $\widehat{\mu}$, and second to improve the estimation of $\sigma^2$.

\begin{figure}[htbp]
\hspace*{-6mm}
\begin{center}
\hspace*{-5mm}
\includegraphics[scale=.58]{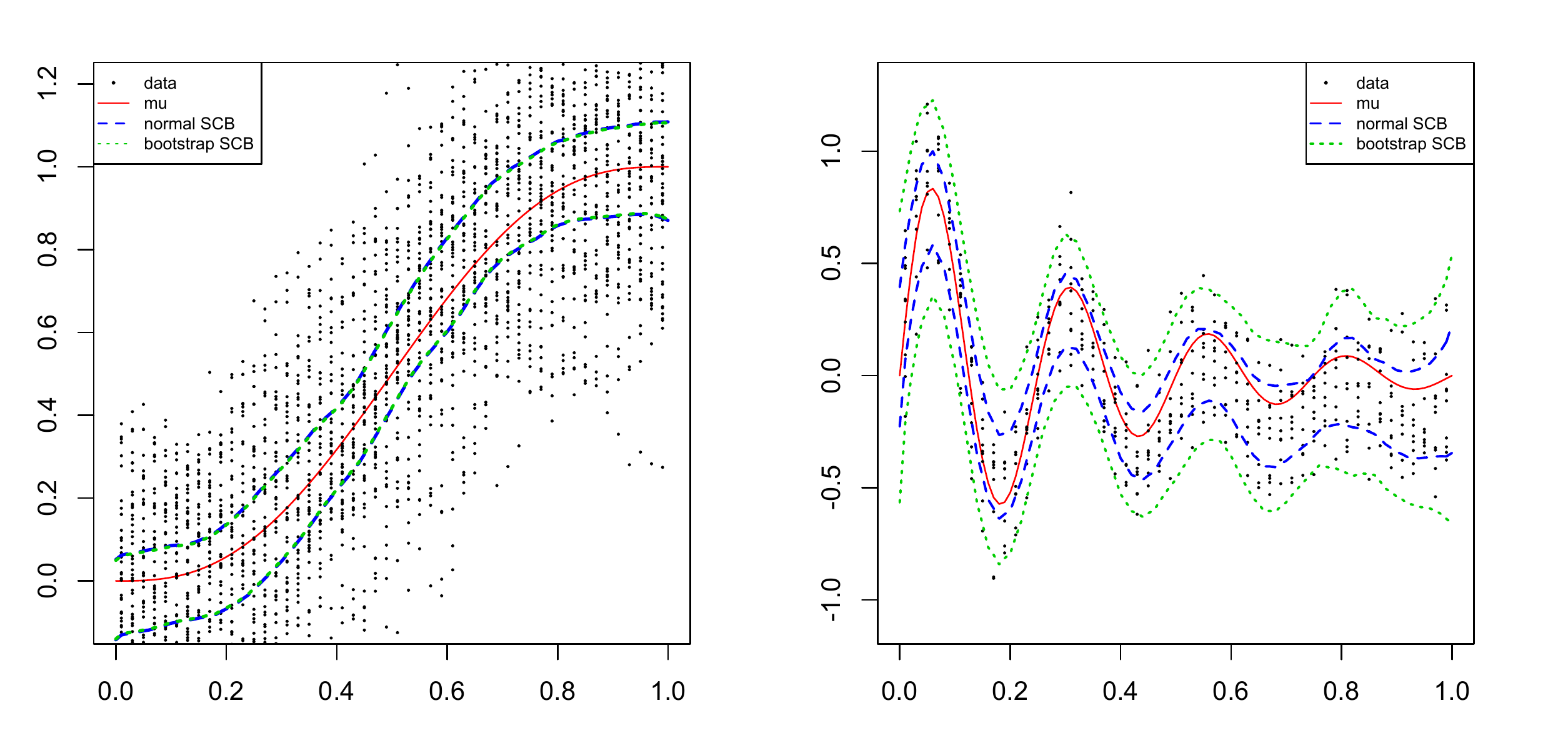} 
\end{center}
\vspace*{-6mm}
\caption{\it SCB for the regression function $\mu$. 
Left panel: model \eqref{model1} with $n=p=50$ and $h = 0.035$. 
The normal and bootstrap bands are identical and achieve the target coverage level 95\%.    
Right panel: model \eqref{model2} with $n=10$, $p=50$, and $h=0.05$. The bootstrap bands are wider than the normal ones ($c_\gamma= 6.50$ vs $c_\gamma=3.28$ in \eqref{SCB})  and have nearly nominal coverage  (94.1\% vs 71.3\%). }
\vspace*{3mm}
\label{figure1} 
\end{figure}


\pagebreak

To conclude this section, we comment briefly on additional simulations not displayed here. 
When simulating model \eqref{model1} with different correlation levels, replacing the parameter $\nu=0.9$ in the covariance $R$ by $\nu=0.7$ or $0.5$, the coverage levels are close to nominal for the two SCB procedures and as expected, the threshold $c_\gamma$ increases as the amount of correlation $\nu$ decreases. Also, when crossing the functions $\mu$ and $R$ of \eqref{model1} and \eqref{model2} in new simulations,  we are confirmed in the idea that the coverage level depends mostly on the (negligibility of) the squared bias to variance ratio and on the (non-)normality of the estimator. \\


\par
\noindent {\bf 5.2 Goodness-of-fit: comparison between the SCB test and a Pseudo-Likelihood Ratio Test}  

This section assesses numerically the statistical significance and the power of the goodness-of-fit test of Section 4.2, 
 referred to as the SCB test henceforth. We use the Pseudo-Likelihood Ratio Test (PLRT) of Azzalini and Bowman (1993) 
as a benchmark for comparison because of its generality and simplicity  of implementation in model \eqref{data model}.
Hereafter we proceed to describe the simulation model, the implementation of the tests, and the experimental results.

The model under study is 
\begin{equation}\label{model3}
\left\{ \begin{array}{l}
Y_{ij} =  \mu(x_j) + Z_i(x_j)  , \quad 1\leq i \leq n , 1 \leq j \leq p, \\
x_j=(j-0.5)/p, \\
H_0: \mu(x)= x, \\
H_n:  \mu(x)= x + n^{-1/2}\log(n) g(x) \\
\textrm{ with } g\in C^2([0,1]), \quad
g(x) = 0 \textrm{ for }  x\in [0, 0.4] \cup [0.6,1] ,\\
 g(x)\in \pi_5  \textrm{ for }  x\in ( 0.4, 0.45] , \quad    g(x)\in \pi_5   \textrm{ for }  x\in ( 0.55, 0.6] , \\ 
g(x)= 0.2 \exp(-(x-0.5)^2)   \textrm{ for }  x\in ( 0.45, 0.55], \\
Z_i  \stackrel{iid}{\sim} Z = \mathcal{G}(0,R) \: \textrm{with } R(x,x') = (0.25)^2 \exp( 20 \log(0.9) | x-x'|),
\end{array} \right.
\end{equation}
where $\pi_k$ denotes the space of polynomials of degree at most k. 
(The covariance structure of the data is the same as in model \eqref{model1}.)  
The candidate model for $\mu$  is $ \mathcal{M}=  \pi_1$, the space of linear functions. 
The local alternatives $H_n$ are obtained by adding a scaled bump function $g$ to $\mu_0(x)=x$,  
which produces a local nonlinearity on $[0.4, 0.6]$.
On account of Section 4.2 we can expect the SCB test  to detect the nonlinearity in $\mu$ while the PLRT might very well miss it.  
\pagebreak 

The simulations were realized similarly to  Section 5.1. 
For different values of $(n,p,h)$, model \eqref{model3} was simulated 50,000 times 
under  $H_0$ and $H_n$, the goodness-of-fit tests were implemented and their type I\&II error rates measured.    
The SCB test required to estimate the covariance function of the test statistic $\sqrt{n} r$ defined by \eqref{residual process} and the threshold $c_\alpha$ of Corollary \ref{test consistency}. The covariance was estimated by  shrinking the empirical covariance of the data 
with the R package \verb@corpcor@ and 
plugging the shrinkage matrix into  \eqref{cov r finite sample} in place of $\boldsymbol{\Sigma} $. 
(Observe that the error covariance matrix $\mathbf{V}$ is zero in the absence  of measurement errors in \eqref{model3}.)  The threshold $c_\alpha$ was estimated as in Section 4.2, using an equispaced grid of size 100 to simulate realizations of the Gaussian process $\mathcal{G}(0,\widehat{\rho}_\Gamma)$ conditional on the correlation estimator $\widehat{\rho}_\Gamma$.  
The PLRT of Azzalini and Bowman (1993) was implemented with standard R packages. 
We give here a short description of this procedure in the context of \eqref{model3}. 
The test starts by fitting a regression line $\widehat{\mu}_{LS}$ and a local linear estimator $ \widehat{\mu}$  
to the averaged data $(x_j,\overline{Y_j})$. It then computes the test statistic $F= RSS_0/RSS_1-1$, where 
$RSS_0$ and $RSS_1$ are the residual sums of squares of $\widehat{\mu}_{LS}$ and $\widehat{\mu}$, respectively. 
Denoting by $F_{obs}$  the observed value of $F$ and putting $\overline{Z}_{p}=(\overline{Z}(x_1),\ldots,\overline{Z}(x_p))^{\top}$,  the $p$-value $\mathbb{P}(F \ge F_{obs}|H_0)$ can be written as $P(  \overline{Z}_{p}\hspace*{-.5mm}^{\top} \mathbf{A} \overline{Z}_{p} >0 )$ for some $p\times p$ symmetric matrix $\mathbf{A}$ depending on $F_{obs}$ and the smoothing matrices of  $\widehat{\mu}_{LS}$ and $\widehat{\mu}$. 
The distribution of $ \overline{Z}_{p}\hspace*{-.5mm}^{\top} \mathbf{A} \overline{Z}_{p}$ is well approximated by an $a\chi^2_b + c$ distribution, where $a,b,c$ depend on $\mathbf{A}$ and $\boldsymbol{\Sigma}$ and are obtained by matching the first three cumulants of the two distributions.  The $p$-value then obtains as $1- P(\chi^2_b \leq -c/a)$. 
Returning to our simulations,  the PLRT required to estimate the unknown covariance    
 $\boldsymbol{\Sigma} =(\sigma^2 \rho^{|j-k|})$, with $\rho= (0.9)^{20/p}$ and $\sigma = 0.25$.  
The estimation was done either with the empirical covariance of the data, 
either by correctly assuming an AR(1) model for the discretized process $(Z(x_1),\ldots,Z(x_p))$
and estimating $\sigma^2$ and $\rho$ through standard repeated measurements techniques 
(e.g. Hart and Wehrly, 1986). 
To  assess the influence of covariance estimation, the PLRT was also implemented with $  \boldsymbol{\Sigma}$ known.

\begin{table}[htdp]
\begin{center}
\begin{tabular}{ c c  c  | c | c c c}
& && \textrm{SCB} & \multicolumn{3}{c}{PLRT}  \\
n& p & h  &Cov. nonpar. & Cov. nonpar.  &Cov. par. &Cov.  known \\
\hline
10& 10 & 0.17 & 0.038 &  0.093 &0.065  &0.051  \\
10& 10 & 0.25 & 0.047 & 0.089 &0.060 &0.052 \\
20 & 20 &0.08 &  0.047 &0.063&0.062&0.050 \\
20 & 20 & 0.15& 0.074& 0.061 & 0.063&0.051 \\ 
50 & 50 & 0.035 &   0.053  &0.049& 0.062&0.050    \\
50 & 50 & 0.05 & 0.061 & 0.050 & 0.063&0.050 \\
100 & 100 & 0.02 &  0.053  &0.047&0.064  &0.050 \\
100 & 100 & 0.05 &  0.063 &0.046 &0.065&0.052  \\
\end{tabular}
\end{center}
\caption{{\it Type I error rates in testing for linearity  in model \eqref{model3} 
at the significance level $\alpha=5\%$. For each $(n,p,h)$ and each test, 50,000 simulations were run. 
In the PLRT procedure, the covariance structure of the data was either estimated nonparametrically, 
parametrically, or known.}}
\label{typeI error rate}
\end{table}

Table \ref{typeI error rate} displays the type I error rates over the simulations. 
For the PLRT procedure, when $  \boldsymbol{\Sigma}$  is known, this rate is very near the significance level $\alpha=5\%$ as expected.  When $  \boldsymbol{\Sigma}$ is estimated parametrically the  rate is slightly excessive, around 6\%. 
In the case of a nonparametric estimation of $  \boldsymbol{\Sigma}$, the PLRT is clearly not accurate for a small sample size $n=p=10$, slightly off for $n=p=20$, and  it works fine for $n=p=50$ and $n=p=100$. 
In comparison, the  SCB test  is overall less accurate regarding the target level $\alpha = 5\%$ except for small sample sizes where it seems more robust. It should be noted however that  for each $(n,p),$ there is at least one value $h$ yielding nearly nominal coverage.

\begin{table}[htbp]
\begin{center}
\begin{tabular}{ c c  c  | c | c c c}
& && \textrm{SCB } & \multicolumn{3}{c}{PLRT}  \\
n& p & h  &Cov. nonpar. & Cov. nonpar.  &Cov. par. &Cov.  known \\
\hline
10& 10 & 0.17 & 0.512 &  0.052 & 0.025 & 0.015 \\
10& 10 & 0.25 & 0.459 & 0.031 &0.012 &0.008 \\
20 & 20 &0.08 &  0.807  & 0.230 & 0.238 &  0.190 \\
20 & 20 & 0.15&  0.664 & 0.061 & 0.059  &   0.040 \\ 
50 & 50 & 0.035 & 0.993   & 0.454 & 0.548  & 0.438    \\
50 & 50 & 0.05 & 0.995 & 0.232 & 0.307 & 0.232 \\
100 & 100 & 0.02 & 1.000   & 0.990 & 0.996 & 0.994 \\
100 & 100 & 0.05 & 1.000  & 0.390 & 0.516& 0.423   
\end{tabular}
\end{center}
\caption{{\it Statistical power of the the SCB and PLRT procedures in model \eqref{model3}. 
The nominal significance level was $\alpha=5\%$ and 50,000 simulations were executed for each $(n,p,h)$ and 
each procedure. For the PLRT procedure, the covariance structure of the data is either estimated 
nonparametrically, parametrically, or known.}}
\label{power lin test}
\end{table}%

\begin{figure}[htbp]
\hspace*{-6mm}
\begin{center}
\hspace*{-5mm}
\includegraphics[scale=.45]{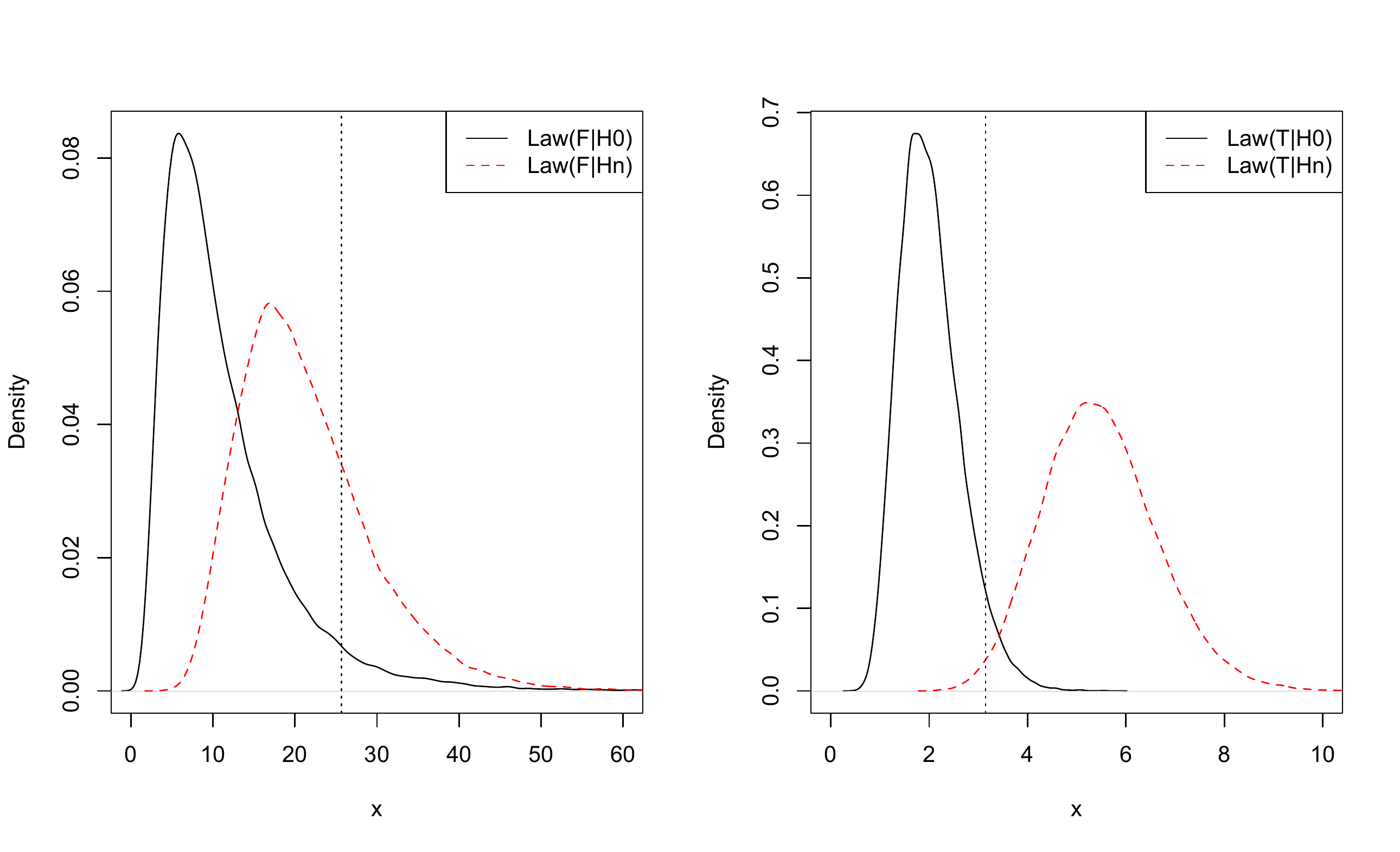} 
\end{center}
\vspace*{-6mm}
\caption{\it Density curves for the PLRT statistic $F$ (left panel) and the sup norm-based statistic $T$ (right panel)  
under $H_0$ and $H_n$ in model \eqref{model3} with $n=p=50$ and $h=0.05$. The vertical lines indicate the critical points for the tests at the level $\alpha = 5\%$. The associated statistical power is 23.6\% for the PLRT and 98.9\% for the SCB test 
(approximated by 23.2\% and 99.5\% in the simulations of Table \ref{power lin test}).}
\vspace*{3mm}
\label{PLRT_SCB_H0_Hn} 
\end{figure}

Looking at Table \ref{power lin test}, it appears that the SCB test has a much larger statistical power than the PLRT. 
Across the simulations, the average power is $80\%$ for the SCB test versus 
about $35\%$ for the PLRT. Besides the 
power does not go below $45\%$ for the 
SCB test while it can be as low as $1\%$--$5\%$ for the PLRT for small samples. In other simulations not displayed here, the superiority of the SCB test gets even larger if the bump function $n^{-1/2}\log(n) g(x)$ in \eqref{model3} is replaced by a smaller bump $n^{-1/2}\log\log(n) g(x)$. Heuristically, the low power of the PLRT can be attributed to the fact that since 
$H_n$ is local in nature and $F$ is based on a euclidean norm, the local discrepancy between $\widehat{\mu}$ and $\widehat{\mu}_{LS}$ at the bump is masked by their global agreement on $[0,0.4] \cup [0.6,1]$. Put differently, 
there is no clear-cut difference between the distribution of $F$ under $H_0$ and under $H_n$ until  $\| \mu_0 -\mu_n \|_{L_2}$ becomes 
 ``large" enough, 
  for large $n,p$ and small $h$. See Figure \ref{PLRT_SCB_H0_Hn} (the densities have been obtained by numerical simulation). 
 Note that analytic power calculations can be obtained for the PLRT via saddlepoint approximations to noncentral $F$ distributions (see e.g. Butler and Paolella, 2002).

\bigskip



\setcounter{chapter}{6}
\setcounter{equation}{0} 
\noindent {\bf 6. Illustration with a speech data set}

\par

In this section we look into a speech data set studied by Hastie et al. (2009) and  
available on the web at  \verb+http://www-stat.stanford.edu~tibs/ElemStatLearn/+.
The data consist in 4509 log-periodograms obtained at 256 equidistant frequencies in the range 0-8kHz. 
Each  discretized curve corresponds to one of five phonemes coded as 'aa' as the vowel in 'dark' (695 curves); 'ao' as  the first vowel in 'water' (1022 curves);   'dcl' as in 'dark' (757 curves); 'iy' as the vowel in 'she' (1163 curves); and 'sh' as in 'she' (872 curves). 
For simplicity of notation we rewrite the observation points  in the frequency domain as $1,\ldots,256$. 

To illustrate the possible uses of SCB techniques,  we present three inference procedures relevant to the statistical analysis  of our 
data set.

\medskip
\noindent{\bf 6.1 Band estimation of regression curves.}

\begin{figure}[htbp]
\vspace*{-2mm}
\begin{center}
\includegraphics[scale=.45]{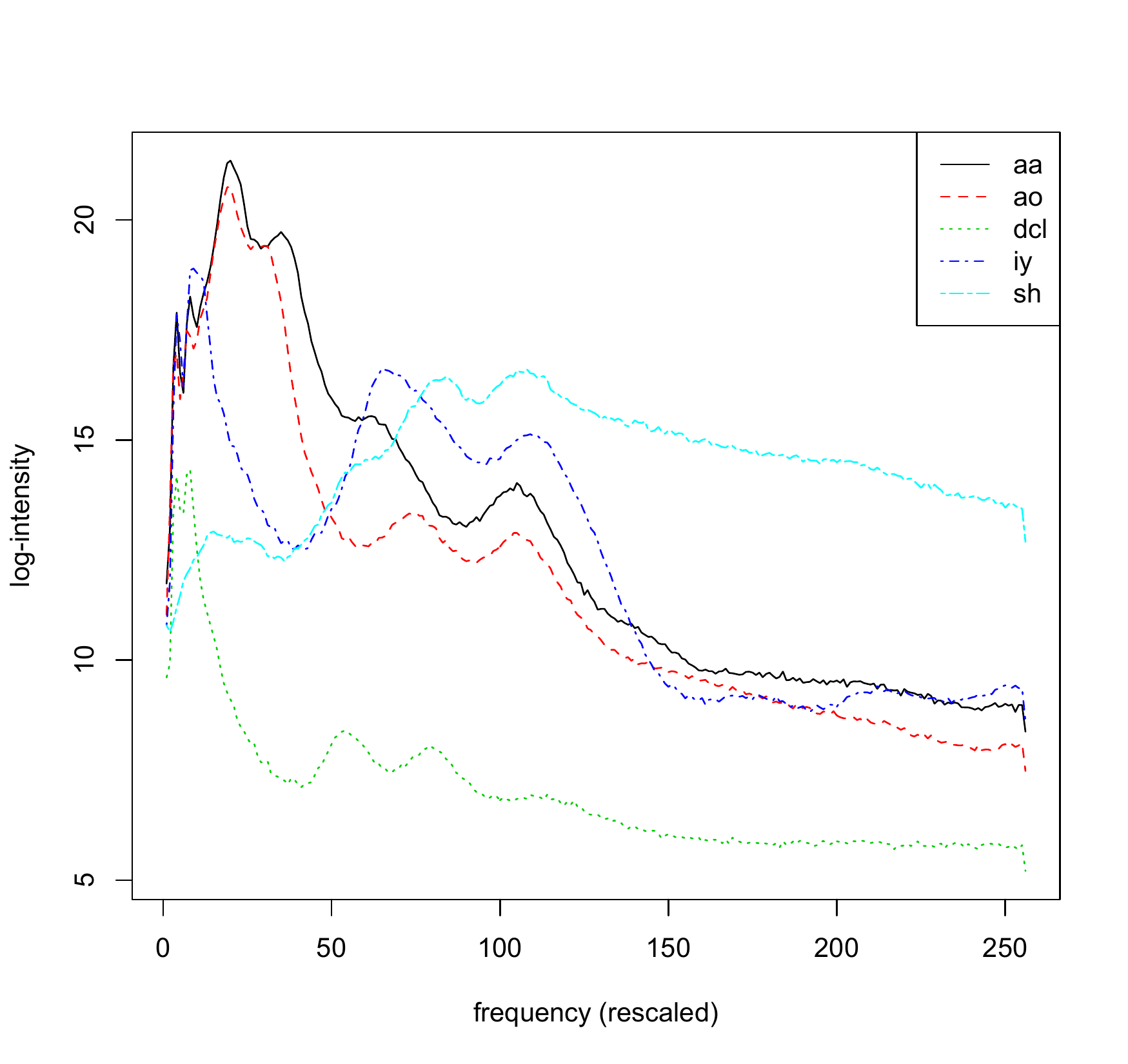} 
\end{center}
\vspace*{-6mm}
\caption{\it Average log-periodograms. 
For each phoneme, the roughness in the individual curves is smoothed out by averaging
but the peaks and valleys are conserved.}
\vspace*{3mm}
\label{speech data average curves} 
\end{figure}

We apply the SCB procedure of Section 4.1 to the mean regression curves for each phoneme. 
Prior to inferring the mean regression curve, it is worth examining  how this mean curve relates to the individual ones.
Indeed it may very well be that the mean curve does not resembles any single curve at all. In our case, the roughness in the
log-periodograms is smoothed out by averaging over the large sample available. 
However some salient features in the individual curves such as peaks and valleys are recovered after averaging, due to the remarkable 
fact that these features are present in almost all the curves at approximately the same locations (see Figure \ref{speech data average curves} and Section 5.3). We observe that  smoothing in the frequency domain seems necessary to make the curves more readily analysable and interpretable.  
The general allure of  the individual smoothed curves (e.g. frequency subdomains where the log-intensity is approximately monotone or linear) is also conserved through the averaging.

\begin{figure}[htbp]
\vspace*{-6mm}
\begin{center}
\hspace*{-5mm}
\includegraphics[scale=.45]{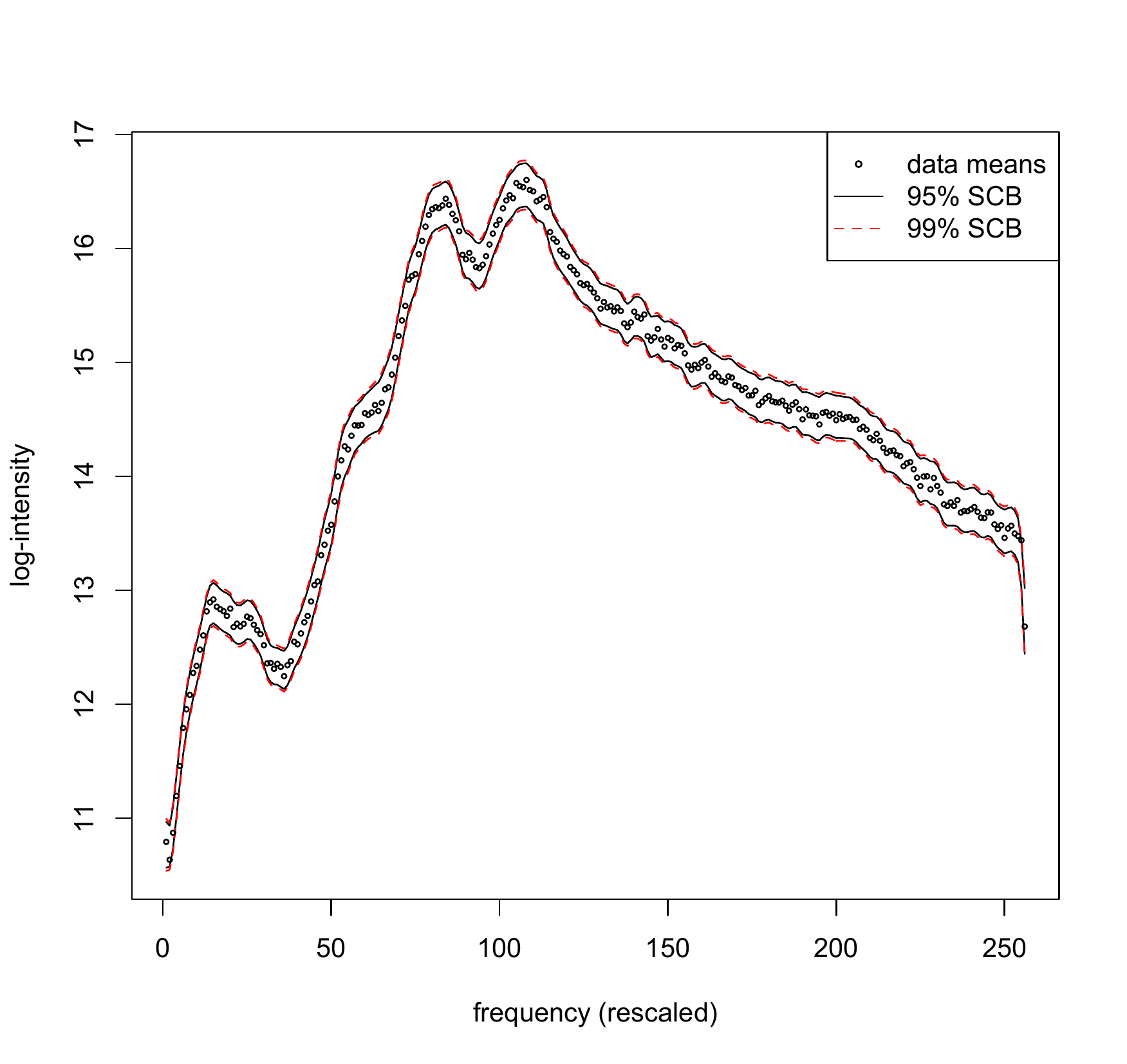} 
\end{center}
\vspace*{-6mm}
\caption{\it SCB of levels 95\% and 99\% for the regression curve of the phoneme 'sh'. Due to the large sample size ($n=872$), the bands have a 
small amplitude allowing to confirm the remarkable features in the regression curve (existence and location of local extrema, monotonicity patterns, etc.). Note that the bands are almost identical at the two confidence levels.}
\vspace*{3mm}
\label{SCB SH} 
\end{figure}

The average log-periodograms are displayed in Figure \ref{speech data average curves}. 
For each phoneme, the empirical standard deviation curve varies between 1 and 3 log-intensity units, which represents a fraction of the range of the average log-periodogram varying between 10\% and 21\% for the phonemes 'aa' and 'ao' (indicating a low variability in the data), 15\% and 30\% for 'dcl' and 'iy' (low to moderate variability), and between 28\% and 35\%  for 'sh' (moderate to large variability). 
For brevity we only show the SCB for the phoneme 'sh' based on the 872 available curves.   
The SCB is built at the levels 95\% and 99\%, using  a local linear estimator with a (truncated) Gaussian kernel (R package \verb@locpoly@) 
and the bandwidth $h=0.94$ that minimizes the leave-one-curve-out cross validation score (see e.g. Hart and Wehrly, 1993).

\medskip
\noindent{\bf 6.2 Comparison of regression curves.}

After building SCB for a single regression curve, we turn to another important inference task which is the comparison of two mean curves. 
Figure \ref{speech data average curves} indicates quite a number of similarities between the regression curves for the phonemes 'aa' and 'ao'. 
A formal inference tool that could confirm or infirm the hypothesis of equality between the two curves would indeed be desirable.   
Such a procedure can be achieved simply by following the method of Section 4.1: 
(i) for each phoneme 'aa' and 'ao', build the corresponding estimator $\widehat{\mu}$ and its estimated covariance $\widehat{R}/n$; 
(ii)  estimate the difference in the regressions $(\mu_{aa}-\mu_{ao})$ by $(\widehat{\mu}_{aa}- \widehat{\mu}_{ao})$ 
whose estimated covariance is $\widehat{R}_{aa-ao}= (\widehat{R}_{aa}/n_{aa} + \widehat{R}_{ao}/n_{ao} ) $ 
(observe that $\widehat{\mu}_{aa} $ and $ \widehat{\mu}_{ao}$ are independent); 
 (iii) denoting by $ \widehat{\sigma}_{aa-ao}$ and $\widehat{\rho}_{aa-ao}$ the standard deviation and correlation functions associated to $\widehat{R}_{aa-ao}$, 
 obtain numerically the distribution of $\| \mathcal{G}(0,\widehat{\rho}_{aa-ao} )\|_\infty $ and  for a given significance level $\alpha$,   
 use the relevant quantile $c_{\alpha} $ of this distribution to build the SCB 
 $\big\{ \big[ (\widehat{\mu}_{aa}- \widehat{\mu}_{ao})(x) \pm c_\alpha \widehat{\sigma}_{aa-ao}(x) \big]: 0\le x \le 256 \big\}$ of level $1-\alpha$ 
 for $(\mu_{aa}-\mu_{ao})$; (v) reject $H_0:\mu_{aa}=\mu_{ao}$ if the horizontal line is not within the bands or equivalently, if $\widehat{\mu}_{aa}$ is not within the bands centered on $\widehat{\mu}_{ao}$. By implementing this procedure with any reasonable bandwidth, $H_0$ is rejected at any significance level ($p$-value $<10^{-16}$).

\begin{figure}[h]
\hspace*{-6mm}
\begin{center}
\hspace*{-5mm}
\includegraphics[scale=.45]{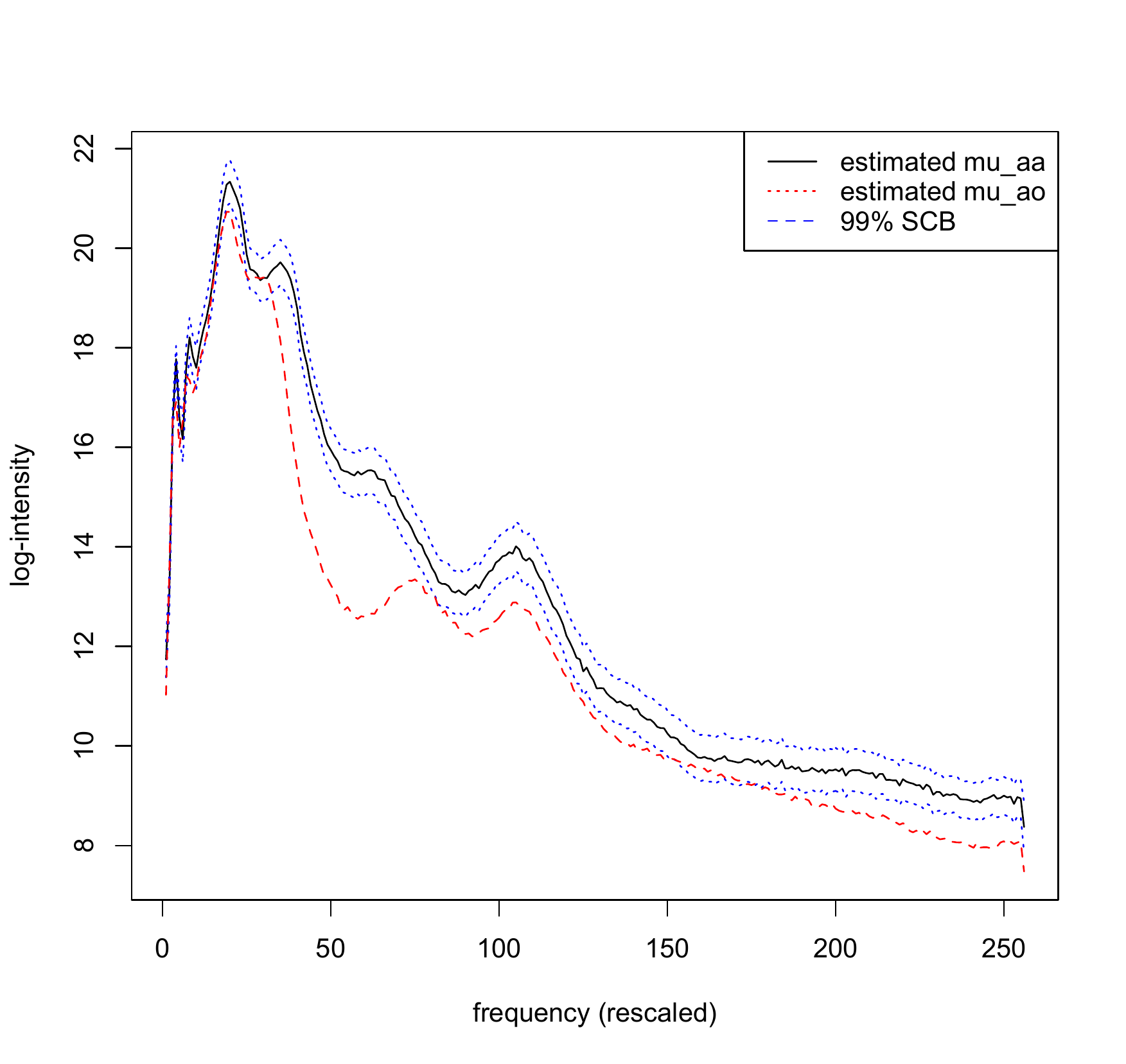} 
\end{center}
\vspace*{-6mm}
\caption{\it Test for equality of the regression curves $\mu_{aa}$ and $\mu_{ao}$. 
SCB of level 99\% are plotted around the estimate $\widehat{\mu}_{aa}$. Since $\widehat{\mu}_{ao}$ is not within the bands, 
the hypothesis $H_0: \mu_{aa}=\mu_{ao}$ can be rejected at the significance level $\alpha=1\%$ (in fact, at any $\alpha$).}
\vspace*{3mm}
\label{SCB SH} 
\end{figure}

\medskip
\noindent{\bf 6.3 Prediction of individual curves.}

The ability to predict new curves and to assess their range of variation proves useful in various situations, e.g. in voice recognition where the goal is to identify the speaker. In the present data set where only a few curves are available for each subject, we study prediction by 
randomly splitting the available curves for each phoneme into a training set and a test set of equal sizes. 
Prediction bands are built from the training set as in Section 4.1 (omitting of course the factor $\sqrt{n}$ in \eqref{SCB} since the goal here is prediction and not regression estimation) and their coverage levels, i.e. the proportions of curves in the test set contained within the bands, are observed in function of the amount of smoothing applied to the data.  Ten fixed bandwidths $h=1,\ldots,10,$ are considered as well as   
a data-driven bandwidth obtained by splitting the training set in half  and selecting the bandwidth $\widehat{h}\in \{ 1,\ldots,10\}$ that gives the closest coverage to the target level 95\% or 99\% for the other half of the training set.  
 For each phoneme the random split is repeated 50 times. The mean coverage levels are reported in Table 
 \ref{prediction scb} for a subset of values of $h$.

Table  \ref{prediction scb} indicates that the coverage levels are very close to nominal as soon as the bandwidth is large enough (and in particular for the data-driven bandwidth $\widehat{h}$) except for the phoneme 'dcl'. These results can be explained by the facts that (i) 
a minimal amount of smoothing is needed to attenuate the erratic, spikey behavior in the raw data curves and make them more predictable; (ii) the distributions of the data curves appear  approximately Gaussian for all phonemes except for 'dcl' which displays strong non-normality. (Our diagnostics for normality were established by performing a functional principal components analysis (PCA) for each set of curves, inspecting visually the plots of the scores along the first few components, and running  Shapiro-Wilks tests on the scores.)

\begin{table}[htbp]
\begin{center}
\begin{tabular}{ c |     c c c c c c  }
$h$ & 1 &    2 &    5  &   8& 10 & $\widehat{h}$ \\ 
\hline
aa &  0.927& 0.938    &0.944 &   0.943&   0.941& 0.945 \\
ao&  0.924 & 0.940&   0.942&  0.947&   0.947 &0.948 \\
dcl & 0.888 & 0.882   &0.884&   0.886&   0.885& 0.890 \\
iy & 0.919 & 0.936    & 0.945&   0.943&   0.944& 0.946 \\
sh & 0.922& 0.941   & 0.949&   0.951&   0.951& 0.952 \\
\hline
aa& 0.985 &0.984&  0.984 &  0.983 &  0.984& 0.987 \\
ao &0.977 &0.984&     0.988&  0.988  & 0.988& 0.989 \\
dcl & 0.945 &0.933&    0.931&   0.932  & 0.932& 0.945 \\
iy & 0.971 &0.984&   0.991&  0.993&  0.992& 0.991 \\
sh& 0.980 & 0.992&    0.992&   0.991  & 0.992& 0.992
\end{tabular}
\end{center}
\caption{{\it \small Coverage levels for the prediction of new curves with SCB of levels 95\% (5 top rows) and 99\% (5 bottom rows). For each phoneme, the bands were based on a random sample comprising half of the available curves and used to predict the remaining half of the curves. The  random sampling was replicated 50 times. The reported numbers are the mean coverage levels over the replications in function of the bandwidth used. The column $\widehat{h}$ denotes a data-driven bandwidth selection procedure.   
}}
\label{prediction scb}
\end{table}

\vspace*{-1mm}



\setcounter{chapter}{7}
\setcounter{equation}{0} 
\noindent {\bf 7. Discussion}

\par

We have established in this paper a functional asymptotic normality result for nonparametric regression with functional data. 
The result allows to build SCB that prove useful in various statistical tasks such as estimating the regression function, testing the goodness of fit of parametric models, testing the equality of mean curves, and predicting individual curves. 
The SCB procedures are fully nonparametric (regression and covariance estimation) and the required bandwidth selection can be data-driven.

It has been seen that the SCB estimation of the regression $\mu$ yields accurate coverage 
whenever $\mu$ is reasonably smooth and sufficient data are available. 
It produces significantly better results than an initial attempt of the author to extend the SCI of Degras (2008) to full bands via the interpolation arguments of Hall and Titterington (1988). (This approach required the difficult estimation of derivatives of $\mu$, causing visually unattractive confidence bands and low coverage.) 
The present SCB estimation of $\mu$, which relies on a numerical method to compute the threshold $c_\gamma$ in \eqref{SCB}, 
 also improves upon previous attempts to approximate $c_\gamma$ via theoretical formulae such as Borell's inequality (see \eqref{BorellInequality} in Section A.2) which is too conservative, or the limit result of Landau and Shepp (1970) which only depends on the confidence level $1-\gamma$ and not on the limit correlation function $\rho$ of $\widehat{\mu}$ (see Degras, 2009).  
(Indeed a sensible estimate of $c_\gamma$ should depend on $\rho$ since the stronger the correlation structure of a (centered, Gaussian) process, the less likely it is to jump above a given threshold $c>0$.) 

On the basis of our numerical study, the SCB goodness-of-fit test clearly outperforms the PLRT of  Azzalini and Bowman (1993) in detecting local departures of $\mu$ from a linear model while retaining a close-to-nominal significance level. This superiority, due to the use of a supremum norm in the test, can be expected to maintain before other tests based on residual sums of squares or $L_2$ distances. On the other hand, the latter kind of test will probably  do a better job at detecting small but global departures from a parametric model. We remark that 
 beyond curvilinear models, the SCB test for goodness-of-fit can be extended e.g. to nonlinear parametric models  fitted by maximum likelihood.

The application of the SCB method to functional prediction (Section 5.3) relies on the approximate normality of the data. 
If normality  does not hold, one may resort to the bootstrap method proposed in Section 4.1.   
 Another use of SCB with potential interest  resides in the estimation of local extrema of the regression function  $\mu$: because the functional asymptotic normality result of this paper also holds for the estimation of $\mu'$ (a formal proof is obviously beyond our scope here), it is possible to build SCB for $\mu'$ and derive confidence intervals for the location and size of local extrema based on the zero crossings of the bands. 
See Song {\it et al.} (2006) for a related work on microarray data. 
 

We say a word about data-driven bandwidth selection and bias reduction. 
Firstly, the popular leave-one-curve-out cross-validation technique appears well suited to our setup  
because of its practical  efficiency and its optimality properties with functional data (Hart and Wehrly, 1993).  
Since by construction the bandwidth $h_{CV}$ in this method is of order $n^{-1/3}$, 
it suffices to slightly strengthen the condition (A.4) into $n^{1/3}\log (p) =o(p)$ for our results to hold with $h_{CV}$ when $d=1$.
Secondly, our results extend easily to jacknife-type estimators of the form $2 \widehat{\mu}_h - \widehat{\mu}_{h\sqrt{2}}$ 
and to local quadratic estimators, which allows to reduce the bias from order $h^2$ (local linear) to $h^3$,  
assuming 3 bounded derivatives  for $\mu$ in (A.1).

Finally, we mention a possible extension to this work which will be of particular interest for  handling functional time series: 
does the functional asymptotic normality of the estimator still hold in the case of dependent data curves? If so, what is the normalizing rate?  
    \\

\noindent {\large\bf Acknowledgment}

The author thanks Professors Michael L. Stein, Wei Biao Wu, and two referees for their suggestions that led to significant improvements in the paper. \\

\appendix

\setcounter{equation}{0} 
\noindent {\bf A.  Proof of Theorem 1}

The proof of Theorem \ref{main result} consists in: first, checking that the squared bias of the local linear estimator $\widehat{\mu}(x)$ defined in \eqref{ll_est}--\eqref{poids LL dim2} is uniformly negligible before its variance over $[0,1]^d$ as $n\to \infty$; second, establishing the conditions of the functional CLT 10.6 of Pollard (1990), 
which mostly amounts to proving the manageability of the smooth curves $\widehat{\mu}_i(x)=\sum_{j=1}^p W_j(x)Z_i(x_j)$; third, showing that the smoothed error process $ \sum_{j=1}^p W_j(x) \overline{\varepsilon}_j$  goes uniformly to zero in probability at a rate faster than $n^{-1/2}$ for $x\in[0,1]^d$. 
The second and third points will be addressed in Sections A.1 and A.2, respectively. 
Putting these results together directly yields the theorem, given the decomposition 
\begin{equation}\label{decomposition}
\widehat{\mu}(x)-\mu(x) = 
\big( \mathbb{E}(\widehat{\mu}(x))- \mu(x)\big) + 
\sum_{j=1}^p W_j (x) \overline{Z}(x_j) + \sum_{j=1}^p W_j (x)\overline{\varepsilon}_j   \,.
\end{equation}
The proof of the theorem being essentially the same in dimensions $d=1, 2$, we only address the univariate case and will briefly indicate how the arguments extend to the bivariate case. Throughout this section the letter $C$ denotes a generic positive constant not depending on $n,p,$ nor $h$.  The notation $J_x =\{ j : | x_j - x | < h \}$ is used for the set  of indexes $j$ for which $W_j(x)\neq 0$ (recall that $K$ has support $[-1,1]$).
The cardinality $| J_x |$ is of order $ph$ due to (A.3).\\

We address here the issue of bias control. With classical bias results for local linear estimators 
(e.g.  Fan (1992)) and Theorem 1 of Degras (2008), it is easy to see that under (A.1)--(A.3), 
\begin{equation*}
\left\{
\begin{array}{l  }
\vspace*{1mm}\sup_{x\in [0,1]^d}\big| \mathbb{E}(\widehat{\mu}(x)) - \mu(x)\big|^2 = \|Ê\mu '' \|_\infty^2 \,  \mathcal{O}\big(\|h\|^4\big)  \\
\sup_{x\in [0,1]^d} \big| \var(\widehat{\mu}(x)) - n^{-1}R(x,x) \big| = o(n^{-1})
\end{array} \right.
\end{equation*}
as $n,p\to \infty$, $h\to 0$ and $p\prod_{k=1}^d  h_k \to\infty$.    
 This entails the condition $n\|h\|^4 \to 0$ in order to make the first negligible before the second. \\


\par
\noindent {\bf A.1 Manageability}   

Let us write $\phi_{in} (x) = n^{-1/2}\sum_{j=1}^p W_{j}(x) Z_i(x_{j})$ for $ i=1,\ldots,n$ 
and $X_n =  \sum_{i=1}^n   \phi_{in}$.
Our aim here is to show the asymptotic normality of $X_n$ in $C ([0,1])$. To do this, we need to establish the conditions (i)--(v) of the functional CLT 10.6 of Pollard (1990).  We start by defining the  objects relevant to this theorem. Let $\Phi_{ni} = n^{-1/2}( |Z_i(0) | + M_i) $ for $i=1,\ldots,n,$ where the $M_i$ are the 
r.v. appearing in assumption (A.2), and consider the envelop $\Phi_n = (\Phi_{n1},\ldots,\Phi_{nn})$ for the $\phi_{in}$. Also define $\rho_n(x,x')= \big[ \sum_{i=1}^n \mathbb{E}\big(\phi_{in}(x) - \phi_{in}(x') \big)^2 \big]^{1/2}$. \\

Using the fact that the  $Z_i$ are independent and distributed as $Z$ and convergence properties of local linear fits, it appears easily that 
\begin{eqnarray}\label{lim pseudo metric}
\rho_n^2(x,x')  \hspace*{-3mm}&=& \hspace*{-3mm} \mathbb{E}\bigg( \sum_{j=1}^{p} (W_{j}(x)- W_{j}(x')) Z(x_{j}) \bigg)^2 \nonumber \\
&=& \hspace*{-3mm} \sum_{j,k} \Big(W_{j}(x)W_{k}(x) - 2 W_{j}(x)W_{k}(x') + W_{j}(x')W_{k}(x') \Big)
R(x_{j},x_{k}) \nonumber \\
& = &\hspace*{-2mm}   R(x,x) - 2R(x,x') + R(x',x') + o(1)
\end{eqnarray}
as $n\to \infty$, $h\to 0$ and $ph\to \infty$. 
Observe that with the same arguments as above, $\mathbb{E}(X_n(x)X_n(x')) \to R(x,x')$ as $n\to\infty$, $h\to 0$ and $ph\to\infty$, which is condition (ii) of the aforementioned theorem. 
Conditions (iii) and (iv) hold because $\sum_{i=1}^n \mathbb{E} (\Phi_{ni}^2) = \mathbb{E}(|Z(0)| +M)^2 < \infty $ by (A.2) and $\sum_{i=1}^n \mathbb{E} (\Phi_{ni}^2 I\{ \Phi_{ni} > \epsilon \} ) = \mathbb{E} \big((|Z(0)| +M)^2 I\{ (|Z(0)| +M) > \epsilon \sqrt{n}\} \big) \to 0$ as $n\to\infty$ for all $\epsilon >0$. 
Condition (v)  is guaranteed by the uniform convergence in   
 \eqref{lim pseudo metric}  which comes from the continuity of $R$ over $[0,1]^2$ 
 (taking expectations in (A.2)) and from the uniformity of the local linear approximation to continuous functions over  compact domains.  \\

It remains to show the more difficult condition (i), namely the manageability property of the $\phi_{ni},i=1,\ldots,n$ with respect to the envelop $\Phi_n$. 
Given an arbitrary real $\epsilon >0$, 
this amounts to evaluating the smallest number $N(\epsilon)$ such that there exist 
$\tau_1,\ldots,\tau_{N(\epsilon)}\in [0,1]$ verifying
$$ \forall x\in [0,1], \: \exists k\in \{ 1,\ldots, N(\epsilon) \} : \:  \forall i \in \{ 1,\ldots, n\}  , \:  
\big| \phi_{ni}(x) - \phi_{ni}(\tau_k) \big| \leq \epsilon \Phi_{ni}.   $$ 
Note that the packing numbers, euclidean norm and rescaling terminology of Definition 7.9 in Pollard (1990) have been rephrased in terms of covering numbers and $l_\infty$ norm 
after observing that $ \sum_{i=1}^n \alpha_i^2 (\phi_{ni}(x) - \phi_{ni}(x'))^2  \leq \epsilon^2 \sum_{i=1}^n \alpha_i^2 \Phi_{ni}^2 $ 
for all rescaling $(\alpha_1,\ldots,\alpha_n)\in \mathbb{R}^n$ is equivalent to   
$ | \phi_{ni}(x) - \phi_{ni}(x') | \leq \epsilon  \Phi_{ni}  $ for $i =1,\ldots,n$.
Let us fix $\epsilon > 0$ and distinguish two cases according to $h=h(n)$. 

\medskip
$\bullet$ $h^\beta \leq \epsilon$. First write 
\begin{equation}
\big| \phi_{ni}  (x)  -Z_i(x) \big| = \bigg| \sum_{j=1}^p W_j(x) \left( Z_i(x_j)-Z_i(x) \right)\bigg| \leq    CM_i h^\beta \leq  C M_i \epsilon
\end{equation} 
for all $x\in [0,1]$ and all $h\geq 1/(2p\max_{[0,1]}(f))$ (the latter condition ensures well-definiteness of local linear smoothing under the design (A.3)), by using the compacity of the support of $K$, 
the H\"older-continuity assumption (A.2) for $Z_i$, and the trivial fact that $\sum_{j=1}^p | W_j(x)|$ 
is uniformly bounded in $x$ and $n$. 
Again with  (A.2), observe that  $\big| Z_i(x)-Z_i(x')\big| \leq CM_i \epsilon$
as soon as  $|x-x'| \leq \epsilon^{1/\beta} $. 
Conclude that for all $(x,x')$ such that  $|x-x'| \leq \epsilon^{1/\beta} $, we have  
\begin{equation}
\big| \phi_{ni}  (x) -\phi_{ni}  (x')   \big| \leq 3C\epsilon  \, \Phi_{ni}   ,
\end{equation}
which yields a covering number $N(\epsilon)$ of the order of  $\epsilon^{-1/\beta}$. \\

$\bullet$ $h^{\beta} > \epsilon$. In this case one easily sees that 
\begin{equation}\label{manage h geq e}
\begin{split}
\big|   & \phi_{ni}  (x)  - \phi_{ni}  (x')  \big| 
\leq   \sum_{j=1}^p \big| W_j(x) - W_j(x') \big| \:    \Phi_{ni}   .
\end{split}
\end{equation}
We  study the previous increment with the following result.


\begin{lemma}\label{rel w ll}
As $n,p\to\infty$, $h \to 0$ and $ph \to\infty$, 
\begin{subequations}
\begin{equation}\label{weight fun bound}
 \big| W_j(x)\big| = \mathcal{O}\left( \frac{1}{ph} K\left( \frac{x-x_j}{h}\right) \right)
\end{equation}
and
\begin{equation}\label{inc bound}
\big| W_j(x) - W_j(x') \big| = \mathcal{O}\left( \frac{1}{ph} \left( \frac{|x-x'|}{h} \wedge 1 \right) \right)
\end{equation}
\end{subequations} \vskip 1mm \noindent 
uniformly in $ j =1,\ldots, p$ and in $x,x'\in[0,1]$, where $a\wedge b = \min(a,b)$. 
\end{lemma}

\noindent {\it Proof.}\\
Recall that $W_j(x)= \frac{w_j(x)}{\sum_{j=1}^p w_j(x)}$, with the $w_j$ defined in  \eqref{poids LL dim1}. 
In view of  assumption (A.3), it can easily be shown that the functions $s_{l}(x)$ defined in \eqref{poids LL dim1} 
satisfy 
\begin{equation*}
 s_{l}(x) = \int_0^1 \frac{(x-u)^l}{h} K\left(\frac{x-u}{h}\right) f(u)du + \mathcal{O}\left(\frac{h^{l-1}}{p}\right) 
\end{equation*}
for $l=0,1,2,$
uniformly in $x \in [0,1]$ as $n\to\infty$ and $h\to 0$. 

Further, it holds that 
\begin{equation*}
\int_0^1 \frac{(x-u)^l}{h} K\left(\frac{x-u}{h}\right) f(u)du  =   (-1)^l h^l  f(x)\,  \int_{-x/h}^{(1-x)/h} u^l K(u)du + \mathcal{O}\left(h^{l+1}\right)
\end{equation*}
and as a consequence, we have 
\begin{equation}\label{eq_sp_ll}
s_{l}(x)  = (-1)^l h^l  f(x)\,  \int_{-x/h}^{(1-x)/h} u^l K(u)du + o\left(h^l\right)
\end{equation}
 and
\begin{equation}\label{eq_sum_wj}
\begin{split}
& \sum_{j=1}^p w_j(x) = s_{2}(x)s_{0}(x) - s_{1}^2(x) = o\big(h^2\big)+  h^2 f^2(x) \\
& \hskip 5mm \times \bigg( \int_{-x/h}^{(1-x)/h} u^2 K(u)du \int_{-x/h}^{(1-x)/h}  K(u)du - \bigg( \int_{-x/h}^{(1-x)/h} u K(u)du \bigg)^2 \bigg) 
\end{split}
\end{equation}
uniformly in $x \in[0,1]$ as $n\to\infty,h\to 0$ and $ph\to\infty$. 

Now, the integral factor in \eqref{eq_sum_wj} is positive by virtue of the Cauchy-Schwarz inequality in $L_2([0,1]^2)$.  (For $x$ far enough from the boundaries this factor reduces to $\int u^2 K(u)du$ 
by the moment properties of $K$ and the compacity of its support  
$K$ is a symmetric density function supported by $[-1,1]$.) 
Moreover, being a continuous function of $x\in [0,1]$ it remains bounded away from zero and infinity  so that 
 $\sum_{j=1}^p w_j(x)$ is uniformly of order $h^2$. 
Invoking \eqref{eq_sp_ll} and the compact support of $K$, one sees
that $w_j(x)=\frac{1}{ph}K\left( \frac{x-x_j}{h}\right) \big( s_2(x) - (x-x_j) s_1(x) \big) = \mathcal{O}\left(\frac{1}{ph}K\left( \frac{x-x_j}{h}\right) h^2 \right)$.  The two former facts on the numerator and denominator of $W_j(x)$ 
produce \eqref{weight fun bound}. 

It remains to compare $ W_j(x)$ and $W_j(x')$ for arbitrary $j,x,x'$. 
First observe that if either $|x-x_j| \geq h$ or $|x' -x_j| \geq h$, then at least one of these weights is zero in which case \eqref{inc bound} reduces to \eqref{weight fun bound}. We may thus assume that $\max( |x-x_j| , |x'-x_j| )<h $. 
In view of the decomposition
\begin{equation}\label{inc decomposition}
 W_j(x) - W_j(x') = \frac{w_j(x) - w_j(x') }{\sum_{j=1}^p w_j(x)} -  W_j(x')  \:  \frac{\sum_{j=1}^p  ( w_j(x) - w_j(x') ) }{\sum_{j=1}^p w_j(x)} \;,
\end{equation}
\eqref{weight fun bound}, the fact that $\sum_j w_j(x)$ is of order $h^2$ and \eqref{poids LL dim1},  
the comparison of the weights  $ W_j(x)$ and $W_j(x')$ 
boils down to comparing 
$s_{l}(x)$ and $s_{l}(x')$ for $l=0,1,2$. 
 
Basic linear algebra shows that 
\begin{equation}\label{diff_sp_st_ll}
\big|  s_{l}(x) - s_{l}(x')\big| =\mathcal{O}\left( \left( \frac{|x-x'|}{h} \wedge 1 \right) h^{l} \right)
\end{equation}
uniformly in $x$ and $x' $.  We now get from \eqref{eq_sp_ll} and \eqref{diff_sp_st_ll} that 
\begin{align}
& ph\: \Big|  w_j(x)-w_j(x') \Big|  \nonumber \\ 
 & =   \left|  \left(s_2(x)- (x-x_j) s_{1}(x)\right) K\left(  \frac{x-x_j}{ph}\right)  -  \left(s_2(x')-  (x'-x_j) s_{1}(x')\right) K\left(  \frac{x'-x_j}{ph}\right)  \right| \nonumber \\
& \leq  
\left| K\left(  \frac{x-x_j}{ph}\right) - K\left(  \frac{x'-x_j}{ph}\right) \right| \cdot  \big|s_2(x)- (x-x_j) s_{1}(x)\big| \nonumber\\ 
  & \quad +  K\left(  \frac{x'-x_j}{ph}\right) \Big\{  \big| (s_2(x)- s_2(x) \big| + \big| s_{1}(x) (x-x') \big|   + \big| (x-x_j)(s_1(x')- s_{1}(x))\big| \Big\}  \nonumber \\
& =     \left( \frac{|x-x'|}{h} \wedge 1 \right) \cdot 
 \mathcal{O}  \left( h^2\right) .
\end{align}

Finally, putting together the fact that $| J_x |$ and $|J_{x'}|$ are of order $ph$ 
(i.e. the non-null weights entering the sum $\sum_{j=1}^p (w_j(x)- w_j(x'))$ in \eqref{inc decomposition} are in a number of order $ph$), 
\eqref{weight fun bound}, \eqref{eq_sum_wj}, and \eqref{diff_sp_st_ll}, 
one may conclude to \eqref{inc bound} without difficulty. $\square$\\

Now, using Lemma \eqref{rel w ll}  and the fact that $|J_x | $ and $|J_{x'} | $ are of order $ph$, one obtains from  \eqref{manage h geq e} the following bound:
\begin{equation}
\Big|    \phi_{ni}  (x)  - \phi_{ni}  (x')  \Big|  \leq  C \left( \frac{|x-x'|}{h} \wedge 1 \right) \Phi_{ni}.
\end{equation}
Since $h^{\beta} > \epsilon$ by assumption, it follows that for 
any $x,x'$  
such that $|x-x'| \leq \epsilon^{1+1/\beta} /C$, the distance between 
$ \phi_{ni}  (x)$ and $\phi_{ni}  (x')$ ($i=1,\ldots,n$)  
is smaller than $\epsilon \Phi_{ni}  $. Therefore the covering number $N(\epsilon)$ 
is at most of the order of $\epsilon^{-1-1/\beta}$. 

Finally, gathering the cases $h^\beta \leq \epsilon$ 
and $h^\beta > \epsilon$, we see that $N(\epsilon)$ is at most of order 
$\max(  \epsilon^{-1/\beta},\epsilon^{-1-1/\beta}) = \epsilon^{-1-1/\beta}$ for $\epsilon <1$, which guarantees 
 the fact that $\int_0^1 \big( \log N(\epsilon) \big)^{1/2} d\epsilon < \infty$, i.e.
the manageability of the $\phi_{ni}$ with respect to the envelop $\Phi_n$. 
All the conditions of the functional CLT 10.6 of Pollard (1990) are thus met. Applying it, we get that 
$X_n = \sqrt{n} \sum_{j=1}^p W_j \overline{Z}(x_j)$ converges weakly in $C ([0,1])$  
to a Gaussian process with mean zero and covariance $R$, as claimed. 
\begin{remark}\label{extension bivariate}
In the bivariate case ($d=2$), the previous arguments carry over with a few simple modifications. In particular in Lemma  \eqref{rel w ll}, \eqref{weight fun bound} transforms into $|W_j(x)| =  \mathcal{O}\big( (ph_1h_2)^{-1} 
   K \big( h^{-1}(x-x_j)\big) \big) $ and \eqref{inc bound} becomes 
$\big| W_j(x) - W_j(x') \big| = \mathcal{O}\left( (ph_1h_2)^{-1}(\|h^{-1}(x-x') \| \wedge 1) \right) $. 
The manageability property is obtained exactly as when $d=1$, by studying four cases according to the signs of $h_1^\beta - \epsilon$ and $h_2^\beta - \epsilon$.  
The other conditions of the functional CLT come alike. 
\end{remark}


\par
\noindent{\bf A.2 Control of the smoothed error process}

Let us  denote by $W(x)$ the vector of weight functions
$(W_1(x),\ldots, W_p(x))^{\top}$ of the local linear estimator at $x$ 
and by   $\widehat{\varepsilon}(x)$ 
the smoothed error process $\sum_j W_j(x) \overline{\varepsilon}_j$.  We will  
 show that $\sqrt{n} \|Ê\widehat{\varepsilon} \|_\infty $ converges to zero in probability  as $n\to\infty$ 
by applying the well-known Borell's inequality
\begin{equation}\label{BorellInequality}
P \left( \sup_{ t \in T} X (t ) > \lambda \right) 
\leq 2 \exp 
\left( - \frac{1}{ 2\sigma_T^2} \left( \lambda - \mathbb{E}\Big( \sup_{t \in T} X(t)\Big) \right)^2 \right)  
\end{equation}
holding for all centered, continuous Gaussian process $X$ indexed by a set $T$ and for all 
$\lambda > \mathbb{E} (\sup_{t \in T} X (t))$, where $\sigma_T^2 =  \sup_{t\in T} \mathbb{E}(X^2(t)) $ (e.g.  Adler (1990) p. 43). 
In the present context $X= \sqrt{n} \widehat{\varepsilon} $ and $T=[0,1]$.

Before to apply  \eqref{BorellInequality}, we must bound the quantities $\mathbb{E} (\sup_{x \in [0,1]} \sqrt{n}\widehat{\varepsilon}(x)) $ and $n \sup_{x \in [0,1]}  \var  ( \widehat{\varepsilon}(x)  )$. 
For the first quantity, we use the classical entropy bound
\begin{equation}\label{entropy}
 \mathbb{E} \Big( \sup_{t \in T} X (t) \Big) \leq C \int_0^\infty \sqrt{ \log N(\epsilon) } \: d \epsilon
\end{equation}
(see e.g. Adler (1990) p.106) where $C>0$ is a universal constant and $N(\epsilon)$ is the smallest number of balls needed to cover $T$ in the pseudo-metric $d(s,t)= (\mathbb{E} (X(s)-X(t))^2)^{1/2}$. 
Here, with assumption (A.5) on the common covariance matrix $ \mathbf{V}$ of the random vectors 
$(\varepsilon_{i1},\ldots,\varepsilon_{ip})^{\top},i=1,\ldots,n$, we have
\begin{eqnarray*}
d^2(x,x') &=&  \mathbb{E} \bigg( \sqrt{n} \sum_{j=1}^p (W_j(x)-W_j(x')) \overline{\varepsilon}_j \bigg)^2 \\
&=& \big( W(x)-W(x')\big)^{\top} \mathbf{V} \big(   W(x)-W(x')\big) \\
&\leq & \|\mathbf{V}\| \times \|  W(x)-W(x') \|^2 \\
&\leq& C  \|  W(x)-W(x') \|^2 
\end{eqnarray*}
where $\|\mathbf{V}\|$ denotes the largest eigenvalue of $ \mathbf{V}$. 
It follows from Lemma \ref{rel w ll} that
 \begin{equation}\label{bound on pseudo-metric}
d(x,x')  \leq  \frac{C}{\sqrt{ph}} \left(  \left| \frac{   x- x' }{h} \right|  \wedge 1 \right)
\end{equation}
and thus, for all $n\geq 1$ and $\epsilon >0 $, it holds that  
\begin{equation}\label{entropy bound}
 \left\{ 
\begin{array}{lc} \vspace*{1mm}
N(\epsilon) = 1 & \textrm{ if } \epsilon \geq \frac{C}{\sqrt{ph}}\, ,\\
N(\epsilon) \leq \frac{C}{ h \sqrt{ph} \epsilon } & \textrm{ if } \epsilon <\frac{C}{\sqrt{ph}} \,.
\end{array}
\right. 
\end{equation}

Plugging \eqref{entropy bound} in \eqref{entropy} we obtain 
\begin{eqnarray}\label{bound for expected supremum}
\mathbb{E} \bigg( \sup_{x \in [0,1]} \sqrt{n} \widehat{\varepsilon}(x) \bigg) & \leq & C \int_0^{\frac{C}{\sqrt{ph}}} \sqrt{ - \log(p^{1/2}h^{3/2} \epsilon) } \: d\epsilon \nonumber \\
& = & \frac{C}{h\sqrt{ph}}  \int_{\sqrt{\log(C/h)} }^{\infty}  u^2 \exp(-u^2) \,   du \nonumber  \\
&\leq &  \frac{C}{h\sqrt{ph}} \times   h\sqrt{|\log(h)|} = C \sqrt{ \frac{|\log(h)|}{ph}} 
\end{eqnarray}
after using the change of variable $u=\sqrt{ - \log(p^{1/2}h^{3/2} \epsilon) }$, an integration by parts, and the classical tail probability bound $\int_{x}^\infty \phi(t)dt < x^{-1}\phi(x)$ (with $x>0$ and $\phi(t)= (2\pi)^{-1/2} \exp(-t^2/2)$).
Hence it suffices that $h\to 0$ and $\log(h)/ph  \to 0$ as $n\to\infty$ to ensure that $\mathbb{E} (\sup_{x \in [0,1]} \sqrt{n} \widehat{\varepsilon}(x)) \to 0$. After simple algebraic manipulations of the condition $\log(h)/ph  \to 0$
together with the rates $n=o(p^4)$ in (A.4) and  $nh^4\to 0$ in Theorem 1, 
it turns out that this condition is equivalent to $h\to 0$, $n^{1/(4d)}\log(p)=o(p)$ in (A.4), and $(p/\log(p))h\to\infty$ in Theorem 1. 
We thus use the latter conditions which are more explicit than the former.

Turning to the variance of $\widehat{\varepsilon}$, we  utilize again Lemma \ref{rel w ll} and 
(A.5) to  get for all $x\in[0,1]$
\begin{equation}\label{variance bound}
\var\left( \widehat{\varepsilon}(x) \right) = \frac{W (x)^{\top}\mathbf{V}W(x)}{n} \leq \frac{\| \mathbf{V} \| \cdot \| W(x) \|^2}{n} \leq  \frac{C}{nph}.
\end{equation}

Borell's inequality \eqref{BorellInequality} may now be applied  to $X=\sqrt{n}\widehat{\varepsilon}$ with $\lambda$ set to an arbitrary $\epsilon >0$. Under the conditions $h\to 0$ and $ph| \log (h)|  \to \infty$ as $n\to \infty$, 
we deduce from \eqref{bound for expected supremum} and \eqref{variance bound}
that
\begin{equation}\label{conclusion noise control}
P \bigg( \sqrt{n} \sup_{ x \in [0,1]}  \widehat{\varepsilon}(x) > \epsilon \bigg) 
= \mathcal{O}  \left(  \exp \left( - C ph \epsilon^2 \right)  \right),
\end{equation}
which yields the uniform convergence in probability of the smoothed error process $\sqrt{n}   \widehat{\varepsilon}$ 
to zero as requested.

\begin{remark}
 The former arguments extend to the bivariate case simply by replacing $h$ with $h_1 h_2$ and $|(x-x') / h| $ with $\| (x-x)' / h \|$ in \eqref{bound on pseudo-metric}--\eqref{conclusion noise control}. 
 In particular  \eqref{bound on pseudo-metric} extends 
 to the case $d=2$ 
 thanks to Remark  \ref{extension bivariate} and a simple partitioning of $[0,1]^2$, while the other equations come in a straightforward way. 
The conclusion then holds under the conditions $\| h \| \to 0$, $ph_1h_2\to \infty$ and $(ph_1h_2)^{-1}\log(h_1h_2)\to 0$, or equivalently   $\| h \| \to 0$ and $p/\log(p) (h_1h_2) \to \infty$.  
\end{remark}


\bigskip

\noindent{\large\bf References}
\begin{description}
\item
Adler, R. J. (1990). An introduction to continuity, extrema, and related topics for general 
Gaussian processes. IMS, Hayward, CA.

\item
 Azzalini, A. and Bowman, A. (1993).  
On the use of nonparametric regression for checking linear relationships. 
{\it  J. Roy. Statist. Soc. B} {\bf  55},  549--557.

\item
Baraud, Y. (2004). Confidence balls in Gaussian regression.
{\it Ann. Statist.} {\bf  32}, 528--551.

\item Butler, R. and Paolella, M. (2002).
Calculating the density and distribution function for the singly and doubly noncentral $F$. 
Statist.  Comput. {\bf 12}, 9--16. 

\item
Degras, D. (2008). Asymptotics for the nonparametric estimation of the mean function of a 
random process. {\it Statist. Probab. Lett.} {\bf  78}, 2976--2980.  
 
\item  
Degras,  D. (2009). Nonparametric estimation of a trend 
based upon sampled continuous processes.
{\it  C. R. Acad. Sci. Paris, Ser. I} {\bf  347},  191--194. 

\item 
Deheuvels, P. and Mason, D. (2004). 
General asymptotic conÞdence bands based on kernel-type function estimators. 
{\it Statist. Infer. Stochast. Proc.} {\bf  7}, 225--277. 

\item
Eubank, R. L. and Speckman,  P. L. (1993).
Confidence bands in nonparametric regression. {\it  J. Amer. Statist. Assoc.} {\bf  88}, 1287--1301.

\item
Fan, J. (1992). Design-adaptive nonparametric regression. 
{\it J.  Amer. Statist. Assoc.} {\bf  87}, 998-1004. 

\item Ferraty, F., and  Vieu, P. (2006). Nonparametric functional data analysis. Theory and practice. Springer Series in Statistics. Springer, New York. 

\item Geman, S. (1980). A limit theorem for the norm of random matrices.  {\it Ann. Probab.} {\bf 8}, 252--261.

\item
Hall, P. and Titterington, D. M. (1988). On confidence bands in nonparametric 
density estimation and regression.  {\it J. Multivariate Anal.} {\bf  27}, 228--254.  


\item 
H\"ardle, W., and Mammen, E. (1993).  
Comparing nonparametric versus parametric regression fits. 
{\it Ann. Statist.} {\bf  21}, 1926--1947.

\item
Hart, J. D. and Wehrly,  T. E. (1986). Kernel regression estimation using repeated 
measurements data. {\it J.  Amer. Statist. Assoc.} {\bf  81}, 1080--1088.

\item 
Hart, J. D. and Wehrly,  T. E. (1993). Consistency of cross-validation when the data are curves. 
{\it Stoch. Proces. Applic.} {\bf 45}, 351--361. 

\item 
Hastie, T., Tibshirani, R., and Friedman, J. (2009). 
The Elements of statistical learning: data Mining, inference, and prediction. Second Edition. Springer Series in Statistics.  Springer, New York.

\item
Johansen, S. and Johnstone, I. M. (1990).  
Hotelling's theorem on the volume of tubes: some illustrations in simultaneous inference and data analysis. 
{\it Ann. Statist.} {\bf 18}, 652--684.

\item
Landau, H. and Shepp, L. A. (1970). 
On the supremum of a Gaussian process. {\it Sankhy\~a} {\bf  32},  369--378. 

\item
Neumann, M. H. and Polzehl, J. (1998). 
Simultaneous bootstrap confidence bands in nonparametric regression. 
{\it J. Nonparam. Statist. } {\bf  9}, 307--333.

\item
Pollard, D. (1990).
Empirical processes: theory and applications. {\it Region. Conf. Ser.  Probab. Statist., vol. 2.} Institute of Mathematical Statistics, Hayward, CA.

\item Ramsay, J. O., and Silverman, B. W. (2005). 
Functional data analysis. Second edition.  Springer Series in Statistics. Springer, New York.
\item
Robinson, P. M. (1997). 
Large-sample inference for nonparametric regression with dependent errors. 
{\it Ann.  Statist.} {\bf  25},  2054--2083. 

\item Song, P. X.-K., Gao, X., Liu, R. and  Le, W. (2006). Nonparametric inference for local extrema with application to oligonucleotide microarray data in yeast genome.  {\it Biometrics} {\bf  62}, 545--554.

\item
Stute, W. (1997). 
Nonparametric model checks for regression.
 {\it Ann. Statist.} {\bf 25}, 613--641.

\item
Sun, J. and Loader, C. (1994).
Simultaneous confidence bands for linear regression and smoothing. 
{\it Ann. Statist.} {\bf  22}, 1328--1347.

\item 
Wang, J. (2009). Modelling time trend via spline confidence band. 
{Submitted. Available at http://www.math.uic.edu/~wangjing/}

\item
Wang, J. and  Yang, L. (2009).
Polynomial spline confidence bands for regression curves.  
{\it Statist.  Sinica} {\bf  19},  325--342. 

\item
Wu, W. B. and Zhao, Z. (2007). Inference of trends in time series. 
{\it J. Roy. Statist. Soc. Ser. B} {\bf  69}, 391--410.

\item
 Yao, F. (2007). 
 Asymptotic distributions of nonparametric regression estimators for longitudinal or functional data. 
 {\it J. Multivariate Anal.} {\bf  98}, 40--56.

\item
Yu, K. and Jones, M. C. (1998).
Local linear quantile regression.
{\it J.  Amer. Statist. Assoc.} {\bf  93}, 228--237.

\item
Zhao, Z. and Wu, W. B. (2008). 
Confidence bands in nonparametric time series regression. 
{\it Ann. Statist.} {\bf  36}, 1854--1878.

\end{description}

\vskip .65cm
\noindent
University of Chicago\\ Department of Statistics
\vskip 2pt
\noindent
E-mail: degras@galton.uchicago.edu

\end{document}